\documentclass[11pt]{article}
\usepackage{xr-hyper}
\usepackage{xcolor}
\usepackage{graphicx}
\usepackage{macros}
\usepackage{longtable}
\usepackage{rotating}
\usepackage{lastpage}
\usepackage{cleveref}
\usepackage{amsthm}
\usepackage{natbib}
\usepackage{url}
\usepackage{authblk}

\definecolor{webbrown}{rgb}{.6,0,0}
\newtheorem{definition}{Definition}[section]
\newtheorem{lemma}{Lemma}
\date{}

\begin{document}
\raggedbottom
	
	\title{Finding Groups of Cross-Correlated Features in Bi-View Data}
	
	\author{Miheer Dewaskar}
	\affil{Department of Statistical Science, Duke University}
	\author{John Palowitch}
	\affil{Google Research, California}
	\author{Mark He}
	\affil{Columbia University Mailman School of Public Health} 
	\author{Michael I.~Love}
	\affil{Department of Biostatistics and Department of Genetics, University of North Carolina at Chapel Hill}
	\author{Andrew B.~Nobel}
	\affil{Department of Statistics and Operations Research and Department of Biostatistics, University of North Carolina at Chapel Hill}

	\maketitle
	
	\begin{abstract}%
		Datasets in which measurements of two (or more) types are obtained from a common set of samples arise in many scientific applications. 
		A common problem in the exploratory analysis of such data is to identify groups of features  of different data types that are strongly associated. 
		A bimodule is a pair $(A,B)$ of feature sets from two data types such that the aggregate cross-correlation between the features in $A$ and those in $B$ is large. 
		A bimodule $(A,B)$ is \emph{stable} if $A$ coincides with the set of features that have significant aggregate correlation with the features in $B$, and vice-versa. 
		This paper proposes an iterative-testing based bimodule search procedure (BSP) to identify stable bimodules. 
		Compared to existing
methods for detecting cross-correlated
features, BSP was the best at recovering true bimodules with sufficient signal, while limiting
the false discoveries. In addition, we applied BSP to the problem of expression quantitative trait loci (eQTL) analysis using data from the GTEx consortium. BSP identified several thousand SNP-gene bimodules. While many of the individual SNP-gene pairs appearing in the discovered bimodules were identified by standard eQTL methods, the discovered bimodules revealed genomic subnetworks that appeared to be
biologically meaningful and worthy of further scientific investigation.
\end{abstract}


\section{Introduction}
\label{sec:intro}

The continuing development and application of measurement technologies in fields such as
genomics and neuroscience means that researchers are often faced with the task of 
analyzing data sets containing measurements of different types derived from a common set of samples.  
While one may analyze the measurements
arising from different technologies individually, additional and potentially important insights can be gained
from the joint (integrated) analysis of the data sets.  Joint analysis, also called multi-view or multi-modal analysis, 
has received considerable attention in the literature, see \cite{lock2013joint, lahatdatafusion,meng2016dimension, tini2019multi, pucher2019comparison, mccabe2019consistency, vahabi2022unsupervised} 
and the references therein for more details.

In this paper we consider a setting in which two data types, Type 1 and Type 2, 
with numerical features are obtained from a common set of samples. 
We refer to correlations between features of the same type as \emph{intra-correlations}, 
and correlations between features of different types as 
\emph{cross-correlations}, noting that this usage differs from that in time-series analysis.
Our primary focus is the problem of identifying pairs $(A,B)$, where $A$ is a set of features of Type 1 and 
$B$ is a set of features of Type 2, such that the aggregate cross-correlation between features in $A$ and $B$ is large 
(see Figure \ref{fig:bimod-def}).   Moreover, we wish to carry out this
identification in an unsupervised, exploratory setting that does not make use of auxiliary information about the samples 
or the features, and that does not rely on detailed modeling assumptions.

Identifying sets of inter-correlated features within a single data type has been widely studied, 
typically through clustering and related methods.  
By contrast, cross-correlations provide information about interactions between data types.  These interactions 
are of interest in many applications, for example, in studying the relationships between 
genotype and phenotype in genomics (discussed in \Cref{sec:eqtl} below),
temperature and precipitation in climate science (discussed in \Cref{sec:climate}), 
habitation of species and their environment in ecology \citep[see, e.g.,][]{dray2003co}, 
and in identifying brain regions associated with experimental tasks \citep{mcintosh1996spatial} 
in neuroscience (cf.~references in \cite{winkler2020permutation} for further neuroscience applications).

Borrowing from the use in genomics of the term ``module'' to refer
to a set of correlated genes, we call the feature set pairs $(A,B)$ of interest to us as {\em bimodules}.  
The term bimodule has also appeared, with somewhat different meaning, 
in \cite{wu_align_2009}, \cite{patel_analysis_2010}, and \cite{pan_bimodule_2019}. 
Bimodules provide evidence for the coordinated activity of 
features from different data types.  Coordination may arise, for example, from shared function, causal interactions,
or more indirect functional relationships.  
Bimodules can assist in directing downstream analyses, generating new hypotheses, 
and guiding the targeted acquisition and analysis of new data. 
By definition, bimodules capture aggregate behavior, which may be significant when no 
individual pair of features has high cross-correlation, or
when the cross-correlation structure between the feature groups is complex.  
As such, the search for bimodules can leverage low-level or complex signals among individual features to
find higher order structure.

\subsection{Existing Work}

Perhaps the easiest way to identify bimodules is to apply a standard clustering method such as k-means to a joint data matrix 
consisting of standardized features from each of the two data types, 
and treating any cluster with features from both data types as a bimodule.  
While appropriate as a ``first look'', this approach requires the analyst to choose an appropriate number of clusters,
imposes the constraint that a feature can belong to at most one bimodule, and
does not distinguish between cross- and intra-correlations. 

A better approach is to investigate bimodules via the sample cross-correlation network.
The cross-correlation network is a bipartite network with vertices corresponding to 
features of Type 1 and Type 2 such that there is an edge between features of different types with a weight equal to their
correlation, but there are no edges between features of the same type.
The CONDOR \citep{platig2016bipartite} procedure identifies bimodules by applying a community 
detection method to an unweighted bipartite graph obtained by thresholding the weights of this cross-correlation network.
One could, in principle, extend this approach by leveraging other community detection methods.

In seeking to uncover relationships between two data types, a number of methods 
identify sets of latent features that best explain the joint covariation between the two 
data types, often by optimizing an objective over the space spanned by the data features 
\citep[see the survey][]{sankaran2019multitable}. 
Sparse canonical correlation analysis (sCCA) \citep{waaijenborg2008quantifying, witten2009penalized, parkhomenko2009sparse} 
finds pairs of sparse linear combinations of features from the two data types that are maximally correlated. 
One may regard each such canonical covariate pair as a bimodule consisting of the features appearing in the linear combination.

In genomics, methods based on Gaussian graphical models
\citep{cheng2012inferring, cheng2015fast, cheng2016sparse} and penalized multi-task 
regression \citep{chen2012two} have been used to find bimodules in eQTL analysis (see
\Cref{sec:eqtl} below). In the former work, 
the authors fit a sparse graphical model with hidden variables that model interactions between 
groups of genes and groups of SNPs.  In \cite{chen2012two}, the gene and SNP networks derived from 
the respective intra-correlations matrices are used in a penalized regression setting to find a 
network-to-network mapping that is similar in spirit to bimodules.

\begin{figure}[tb]
	\centering
\begin{tikzpicture}[xscale=1.7, yscale=1.5]
	\draw (0,0) rectangle (3,1); 
	\draw (4,0) rectangle (6,1);
	
	\node[left] at (-0.1,0.5) {$X$};
	\node[right] at (6.1,0.5) {$Y$};
	
	\node at (3.5, 0.5) {samples};
	\node [below] at (1.5, 0) {\small Type 1 features};
	\node [below] at (5, 0) {\small Type 2 features};
	\draw[fill=BimodColor] (1.5,0) rectangle (2.5,1);
	\draw[fill=BimodColor] (4.5,0) rectangle (5.5,1);
	
	\node (A) [above] at (2,1) {$A$};
	\node (B) [above] at (5,1) {$B$};
	
\draw  (A) edge[<->, bend left] node[midway, above] { cross-correlated }  (B);
\end{tikzpicture}
\caption[Multi-view data and bimodules]{Illustration of a bimodule $(A,B)$ (shaded) arising in two numeric data matrices $X$ and $Y$ matched by samples (rows).
The columns of $A$ and $B$ need not be contiguous.}
\label{fig:bimod-def}
\end{figure}

\subsection{Our Contributions}
\label{sec:bsp-intro}

In this paper we propose and analyze an exploratory method called BSP (an acronym for Bimodule Search Procedure) that identifies bimodules 
in moderate and high dimensional data sets.  
The BSP method is unsupervised: it does not rely on external or auxiliary information about the data at hand.
BSP differs fundamentally from the existing methods described above, in that  
it does not make use of a detailed statistical model to define bimodules, 
and does not treat the sample cross-correlation network as a 
sufficient statistic for bimodule discovery.
Instead, BSP is based on the notion of a \emph{stable} bimodule, which is introduced and studied below.
A bimodule $(A,B)$ is {\em stable} if $A$ coincides with the features of Type 1 that have significant 
aggregate correlation with the features in $B$, 
and $B$ coincides with the features of Type 2 that have significant 
aggregate correlation with the features in $A$. (A formal definition is given in Section \ref{sec:sample-bimod}.)
BSP employs an iterative multiple testing based procedure to identify stable bimodules.
Stability provides a statistically principled criterion for filtering bimodules having significant 
aggregate cross-correlations. 

To handle large datasets with hundreds of thousands of features, we develop and employ a fast analytical approximation 
to permutation p-values that are used to assess the significance of aggregate cross-correlations between features of different types.  
Importantly, these p-values account for intra-correlations (between features of the same type) when assessing the 
significance of observed cross-correlations.
Additionally, we provide a method to extract statistically motivated essential-edge networks from bimodules, enhancing their interpretability. 
We validate BSP through a comprehensive and carefully constructed simulation framework, and demonstrate its 
practical utility through an extended application to eQTL analysis.

\subsection{Organization of the Paper}

The next section presents the testing-based definition of stable bimodules based on
p-values derived from a permutation null distribution and describes the Bimodule Search Procedure and supplemental details of the methodology.  
\Cref{sec:sim} is devoted to a simulation study that uses a complex model 
to capture some aspects of real bi-view data. Here, we evaluate the performance of BSP and compare it to that of CONDOR, sCCA, and \meqtl.
\Cref{sec:eqtl} describes and evaluates the results of BSP, CONDOR, and \meqtl\ applied to an eQTL dataset from the GTEx consortium. 
In particular, we examine the bimodules produced by BSP using a variety of descriptive and biological metrics, 
including comparisons with, and potential extensions of, standard eQTL analysis.

\section{Stable Bimodules and the Bimodule Search Procedure}
\label{sec:sample-bimod}

This section defines stable bimodules and describes the Bimodule Search Procedure and auxiliary details of our methodology. 
We first describe a permutation null-distribution (\Cref{sec:perm-null-dist-and-pvals}) that will be used to provide a testing-based definition of  
stable bimodules (\Cref{subsec:empirical-bimodules}) and the Bimodule Search Procedure (\Cref{sec:bsp}).
We conclude the section with auxiliary details like fast p-value approximation (\Cref{sec:pval-approx}), post-processing bimodules 
to address overlap (\Cref{sec:post-processing}), a network-based assessment of minimality of bimodules (\Cref{sec:connectivity}), tuning the false-discovery parameter using a half-permutation scheme (\Cref{sec:howtochoosealpha}), and suggestions for a potential practical workflow using our methodology (\Cref{sec:workflow}).

\subsection{Notation}
\label{sec:setup}

Let $\bbX$ be an $n \times p$ matrix containing the data of Type 1, and let $\bbY$ be an $n \times q$ 
matrix containing the data of Type 2.  
The columns of $\bbX$ and $\bbY$ correspond to features of Type 1 and Type 2, respectively.   
The $i$th row of $\bbX$ ($\bbY$) contains measurements of Type 1 (Type 2) on the $i$th sample. 
Features of Type 1 will be indexed by $S = \set{s_1, s_2, \ldots, s_p}$, features of Type 2 by $T = \set{t_1, t_2, \ldots, t_q}$. 
Let $\bbX_s$ be the column of $\bbX$ corresponding to feature $s$, and let $\bbY_t$ be the column of $\bbY$ corresponding to feature $t$. 
For $s \in S$ and $t \in T$ let $r(s,t)$ be the sample correlation between $\bbX_s$ and $\bbY_t$. 
For $A \subseteq S$ and $B \subseteq T$, define the aggregate squared correlation between $A$ and $B$ by
\begin{equation}
\label{eqn:asc}
	r^2(A, B) \doteq  \sum_{s \in A, t \in B} r^2(s, t).
\end{equation}
For singleton sets we will omit brackets, writing $r^2(s, B)$ and $r^2(A, t)$.  

The aggregate correlations $r^2(A, B)$ are an obvious starting point for the identification of bimodules.
Maximization of $r^2(A, B)$ subject to a penalty on the cardinalities of $A$ and $B$ is not computationally 
feasible, would typically involve the introduction of free parameters whose values might
be difficult or time-consuming to specify, and would not account for intra-correlations.
The BSP method attempts to address these issues by taking a different approach that is based on iterative multiple testing  
of the statistics $r^2(s, B)$ and $r^2(A, t)$.  The requisite p-values are described next.

\subsection{Permutation P-values} 
\label{sec:perm-null-dist-and-pvals}

It is assumed in what follows that the data matrices $\bbX$ and $\bbY$ are given and fixed.  In this setting,
we define a permutation null distribution that is obtained by randomly reordering the rows of 
$\bbX$ and, independently, the rows of $\bbY$.

\begin{definition}
\label{def:perm-dist}
Let $P_1, P_2 \in \{0,1\}^{n \times n}$ be chosen independently and uniformly 
from the set of all $n \times n$ permutation matrices.
The \emph{permutation null distribution} of $[\bbX, \bbY]$ is the distribution of the data matrix 
$[\tilde{\bbX}, \tilde{\bbY}] \ \doteq \ [P_1 \, \bbX, P_2 \, \bbY]$. 
Let $\mathbb{P}_\pi$ and $\mathbb{E}_\pi$ denote probability and expectation, respectively, 
under the permutation null. 
\end{definition}

For $s \in S$ and $t \in T$ let $R(s,t)$ be the (random) sample-correlation of 
$\tilde{\bbX}_s$ and $\tilde{\bbY}_t$ under the permutation null $\mathbb{P}_\pi$.
Permutation preserves the sample correlation between 
the features $s_i$ in $S$, and between the features $t_j$ in $T$, but nullifies cross-correlations between $S$ and $T$. 
The values of $R(s,t)$ will tend to be small, and in particular $\mathbb{E}_\pi [ R(s,t) ] = 0$ for each $s \in S$ and $t \in T$
(see \cite{fredPerm} for more details).

\begin{definition}
For $A \subseteq S$ and $B \subseteq T$ define the permutation p-value
   \begin{equation}
        p(A, B) \ \doteq \ \mathbb{P}_\pi ( \, R^2(A, B) \geq r^2(A, B) \, ) .
    \end{equation}
    Here $R^2(A, B) \doteq \sum_{s \in A, t \in B} R^2(s, t)$ is random, while the observed sum of squares $r^2(A, B)$,
    defined as in (\ref{eqn:asc}), is fixed.
    \label{def:pvals}
\end{definition}

Small values of $p(A, B)$ provide evidence against the null hypothesis that the observed cross-correlation between 
the features in $A$ and $B$ could have arisen by chance. 
At the same time, the permutation distribution preserves the correlations between features of the same type.  
{\em Thus, the p-value $p(A, B)$ accounts for the effects of intra-correlations when assessing the significance 
of the aggregate cross-correlation $r^2(A, B)$.}   In practice, we employ an approximation of the permutation p-values $p(A, B)$, which is
described in Section \ref{sec:pval-approx} below.

\subsection{Stable Bimodules}
\label{subsec:empirical-bimodules}

Before describing the bimodule search procedure in the next subsection, we introduce stable bimodules, which are 
the targets of the procedure.  Stable bimodules are defined using the multiple testing procedure of
\cite{benjamini2001} (B-Y). 
Let $p = p_1, \ldots, p_m \in [0,1]$ be the p-values associated with a family of $m$ hypothesis tests, 
with order statistics $p_{(1)} \leq p_{(2)} \ldots \leq p_{(m)}$.
Given a target false discovery rate $\alpha \in (0,1)$, the B-Y procedure 
rejects the hypotheses associated with the p-values $p_{(1)}, \ldots, p_{(k)}$ where 
\begin{equation}
p_{(k)} \ = \ 
\max \left\{ p_{(j)} \, : \, p_{(j)} \, \leq \, \frac{\alpha \, j}{m \, \sum_{i=1}^m i^{-1}} \right\} 
 \ \doteq \ \tau_{\alpha}(p)
\end{equation}
As shown in Theorem 1.3 of \cite{benjamini2001}, regardless of the joint distribution of the p-values in $p$, 
the expected value of the false discovery rate among the rejected hypotheses is at most $\alpha$.  
The value $\tau_{\alpha}(p)$ acts as an adaptive significance threshold.

\begin{definition}{(Stable Bimodule)}
Let $[\bbX, \bbY]$ and $\alpha \in (0,1)$ be given. A pair $(A,B)$ of non-empty sets $A \subseteq S$ and $B \subseteq T$ is a  
\emph{stable bimodule} at level $\alpha$ if
\begin{enumerate}
\vskip.05in
    \item $A = \{ s \in S : p(s, B) \leq \tau_{\alpha}(p_B) \}$ and
    \vskip.05in
    \item $B = \{ t \in T : p(A, t) \leq \tau_{\alpha}(p_A) \}$
\end{enumerate}
where $p_{B} = \{p(s,B)\}_{s \in S}$ and $p_{A} = \{p(A, t) \}_{t \in T}$.
\label{def:bimod}
\end{definition}

A bimodule $(A,B)$ is stable if and only if $A$ is exactly the set of features $s \in S$
for which $r^2(s,B)$ is significant, and $B$ is exactly the set
of features $t \in T$ for which $r^2(A,t)$ is significant.  Significance is assessed via the permutation
null and the B-Y multiple testing procedure.  Note that the empty bimodule $\emptyset \times \emptyset$
satisfies conditions 1 and 2 of the definition, but we reserve the term stable for non-empty bimodules.

When $B$ is fixed, the condition $p(s, B) \leq \tau_{\alpha}(p_{B})$ can be written equivalently as 
$r^2(s,B) \geq \hat{\gamma}(s, B)$ where $\hat{\gamma}(s, B)$ is an adaptive correlation threshold 
depending on $s$ and $p_{B}$.
The latter condition may be satisfied even if the feature $s$ is not significantly correlated with any 
{\em individual} feature in $B$.  Similar remarks apply to $p(A,t)$.
In this way stable bimodules can facilitate the aggregation of small effects across feature pairs.
Importantly, stable bimodules cannot be recovered from the sample 
cross-correlation matrix or the associated cross-correlation network alone. 
This is because the thresholds $\hat{\gamma}(s, B)$ and $\hat{\gamma}(A, t)$ also depend on other aspects of the data,
in particular the intra-correlations of features in $A$ and $B$, through the p-values in $p_A$ and $p_B$.

\subsection{The Bimodule Search Procedure (BSP)} 
\label{sec:bsp}

Let $2^S$ be the family of all subsets of features of Type 1, and define $2^T$ similarly. 
According to Definition \ref{def:bimod}, stable bimodules are exactly the non-empty fixed points of the set map 	
$\Gamma: 2^S \times 2^T \to 2^S \times 2^T$
defined by $\Gamma(A,B) = (A', B')$ where
\[
A'  = \{ s \in S : p(s, B) \leq \tau_{\alpha}(p_{B})\} \ \mbox{ and } \ 
B' = \{ t \in T : p(A',t) \leq \tau_{\alpha}(p_{A'})\}. 
\]
The bimodule search procedure (BSP) finds stable bimodules by repeatedly 
applying the map $\Gamma$ to an initial pair $(A_0, B_0)$.  
As the map $\Gamma$ is deterministic, and the number of feature set pairs is finite, 
repeated application of $\Gamma$ is guaranteed to yield a fixed point or enter a limiting cycle.
BSP outputs non-empty fixed points, which are stable bimodules at level $\alpha$.  

In practice, BSP is run repeatedly, initializing with every pair $(A_0,B_0)$, where 
$A_0 = \{s\}$ is a single feature in $S$ and $B_0$ is the set of features $t \in T$ that are significantly
correlated with $s$, or $B_0 = \{t\}$ is a single feature in $T$ and $A_0$ is the set of features $s \in S$ that are significantly correlated with $t$.
Pseudocode for BSP is given in Algorithm \ref{alg:bsp}.
   
\begin{algorithm}
	\SetKwRepeat{Do}{do}{while}
			
	\KwIn{Data matrices $\bbX$ and $\bbY$, parameter $\alpha \in (0,1)$, and initialization 
	set $(A_0, B_0)$, where $A_0 \subseteq S$ and $B_0 \subseteq T$.}
	\vskip.1in
	\KwOut{A stable bimodule $(A,B)$ at level $\alpha$, if found.}
\vskip.1in
	{\bf initialize:} $A' = A_0$, $B' = B_0$, and $A = B = \emptyset$\;
	\While{$(A', B') \neq (A, B)$}{ \label{line:bsploop}
		$(A, B) \gets (A', B')$\; 
		Compute $p(s,B)$ for each $s \in S$ and let $p_{B} \gets (p(s,B))_{s \in S}$\; 
		$A' \gets \{ s \in S : p(s,B) \leq \tau_{\alpha}(p_{B}) \}$\tcp*[r]{Indices rejected by B-Y}

		Compute $p(A',t)$ for each $t \in T$ and let $p_{A'} \gets (p(A',t))_{t \in T}$\; 
		$B' \gets \{ t \in T : p(A',t) \leq \tau_{\alpha}(p_{A'}) \}$\tcp*[r]{Indices rejected by B-Y}
	} 
	\If{$|A||B| > 0$ and $(A,B)=(A',B')$}{
		\KwRet{$(A,B)$}\;
}
\caption{Bimodule Search Procedure (BSP)} \label{alg:bsp}
\end{algorithm}

To limit computation time and address the possibility of cycles, the BSP is terminated after 20 iterations. 
In our simulations and real-data analyses (discussed below) the 20 iteration limit was rarely reached:
cycles were very rare, and most initial conditions led to empty fixed points.

A likely side effect of any directed search for bimodules $(A,B)$, stable or otherwise, is that the sample
intra-correlations of the features in $A$ and $B$ will be large, often significantly larger
than the intra-correlations of a randomly selected set of features with the same cardinality.  
Failure to account for inflated intra-correlations will lead to underestimates of the standard error
of most test statistics, including the sum of squared correlations used here,
which will in turn lead to anti-conservative (optimistic) assessments of significance, and oversized feature sets.  
As noted above, the permutation distribution leaves intra-correlations
unchanged, while ensuring that cross-correlations are close to zero, and as such, the p-values $p(s,B)$ and 
$p(A,t)$ account for intra-correlations among features in $B$ and $A$, respectively.

The false discovery rate $\alpha \in (0,1)$ used in the B-Y multiple testing procedure is the only free 
parameter of BSP. 
While $\alpha$ controls the false discovery rate of the features $s$ or $t$ selected
at each step of the search procedure, it does not guarantee control of
the false discovery rate of pairs $(s,t)$ within stable bimodules.
In general, BSP will find fewer and smaller bimodules when $\alpha$ is small, and will find more and larger bimodules 
when $\alpha$ is large.  In practice, we employ a permutation based procedure to select $\alpha$ from a fixed grid of values 
based on a half-permutation procedure that is described in \Cref{sec:howtochoosealpha}.

Running BSP with multiple initializations results in a list of stable bimodules.  
In order to limit the number of potentially spurious bimodules, and to limit the number of
bimodules where both $A$ and $B$ are singletons, we remove from the list any bimodule $(A,B)$ for which $p(A,B)$ exceeds the 
Bonferroni threshold $\alpha / |A| |B|$ for singleton bimodules.  
Once this preliminary filtering is complete, two, more critical, post-processing issues remain.

The first issue is overlap.  This arises from the fact that
distinct bimodules can exhibit substantial overlap.  In many applications, e.g., the study of gene regulatory networks, 
overlap of interacting feature sets is the norm, and the ability to capture this overlap is critical for successful exploratory analysis.
Nevertheless, extreme overlap can impede interpretation and downstream analyses.  
We deal with overlap in two ways.  First, our focus on stable bimodules eliminates
most small perturbations from consideration.  
Second, in cases where we find two or more stable bimodules with large overlap, 
we employ a simple post-processing step to identify a representative bimodule from the overlapping group. 
This post-processing step is described in Section \ref{sec:post-processing}.

In practice,
we wish to identify minimal bimodules, those that do not properly contain another bimodule, as
such bimodules are more likely to reveal interpretable interactions between features.  
Checking minimality of a stable bimodule can be difficult, and we rely instead on a 
notion of robust connectivity that leverages the connection between bimodules and the 
 sample cross-correlation network.  Informally,
a robustly connected bimodule cannot be partitioned into two groups without removing one or more high weight connections
between the groups.  See Section \ref{sec:connectivity} for a discussion and more details.

Iterative testing procedures have been applied in single data-type settings for community 
detection in unweighted \citep{wilson2014testing} and weighted \citep{palowitch2016continuous} networks, differential correlation mining \citep{bodwin2018testing}, 
and association mining for binary data \citep{mosso2021latent}.  
In these papers a stable set of significant nodes or features is identified through the 
iterative application of multiple testing.  However, the hypotheses of interest and the associated test statistics all 
differ substantially from the setting here.  Moreover, the p-values
in these papers were derived from asymptotic normal and binomial approximations rather than the more complicated permutation moment
fitting used here.

\subsection{Approximation of Permutation P-Values}
\label{sec:pval-approx}

The BSP method relies on evaluating the permutation p-values $p(s, B)$ and $p(A, t)$.
There is no closed form expression for these p-values and direct evaluation 
via permutation is computationally prohibitive, since in each iteration BSP requires the evaluation of
$|S| + |T|$ such p-values. 
As an alternative, we adapt ideas from \cite{zhou2019marker} to approximate the permutation 
p-value $p(A,t)$ (or $p(s, B)$) using the tails of a location-shifted Gamma distribution that has the same first three moments 
as the sampling distribution of $R^2(A,t)$ under the permutation null. Details now follow.

Let $\tilde{\bbX}$ and $\tilde{\bbY}$ be permuted versions of the observed data matrices, as in Definition \ref{def:perm-dist}.
Given $A \subseteq S$ let $\Sigma_A$ be the $|A|\times|A|$ sample correlation matrix of the columns of
$\tilde{\bbX}$ (or equivalently $\bbX$).  It is shown in \cite{dewaskar2021high} that if the random permutation matrices 
in Definition \ref{def:perm-dist} are replaced by random orthogonal matrices  
that fix the constant vector then, conditional on $\Sigma_A$, for each $t \in T$ the first three moments of $R^2(A,t)$ are given by:
\begin{align*}
	m_1 \doteq \mathbb{E}[ R^2(A,t) | \Sigma_A] &= \frac{|A|}{(n-1)} \\
	m_2 \doteq \mathbb{E}[ R^4(A,t) | \Sigma_A] &= \frac{2\sum_{i=1}^{|A|} \lambda_i^2 + |A|^2}{n^2-1} \\
	m_3 \doteq \mathbb{E}[ R^6(A,t) | \Sigma_A] &= \frac{|A|^3 + 6|A|(\sum_i \lambda_i^2) + 8 \sum_i \lambda_i^3}{(n^2-1)(n+3)},
\end{align*}
where $\{\lambda_i\}_{i=1}^{|A|}$ are the eigenvalues of the intra-correlation matrix $\Sigma_A$.  Earlier work of 
\cite{zhou2019marker} establishes the same relations under the stronger assumption that $\tilde{\bbY}_t$ is multivariate
normal and independent of $\tilde{\bbX}_A$.  
We use the above formulas to approximate the moments of $R^2(A,t)$ under the permutation null of Definition \ref{def:perm-dist},
yielding approximate tail probabilities
\[
p(A, t) = \prob(Y \geq r^2(A,t))
\]
where $Y$ has a location shifted Gamma distribution satisfying $\mathbb{E}[Y] = m_1$, 
$\mathbb{E}[Y^2] = m_2$, and $\mathbb{E}[Y^3] = m_3$.  More general moment formulas in \cite{dewaskar2021high} enable us to 
estimate $p(A,B)$ when neither $A$ nor $B$ is a singleton.

As noted above, the permutation based p-values
employed by BSP explicitly account for correlations
between features of the same type through the eigenvalues of the intra-correlation matrix $\Sigma_A$, 
attenuating the significance of cross-correlations when intra-correlations are high.

\subsection{Post-Processing of Bimodules to Address Overlap}
\label{sec:post-processing}

Let $\cB = (A_1, B_1), \ldots (A_N, B_N)$ be the list of stable bimodules found by BSP after initial filtering
to remove bimodules such that $p(A,B)$ exceeds the Bonferroni threshold $\alpha / |A| |B|$ for singletons.  
In general, the same bimodule may appear multiple times in $\cB$, and bimodules
in $\cB$ may overlap.  To address this, we first assess the effective cardinality $N_e$ of $\cB$, then identify
$N_e$ groups of related bimodules in $\cB$, and finally select a representative bimodule from each group.  
Details are given below.

Let $C_j = A_j \times B_j$ be the set of $(s,t)$ pairs in the $j$th bimodule of $\cB$.  Following \cite{shabalin2009finding} we define the \emph{effective number} of bimodules in $\cB$ as
\begin{align}
	\label{eq:effnum}
	N_e & \doteq \sum_{C \in \cB} 
	\left\{ \frac{1}{|C|} \sum_{(s,t) \in C} \,
	\frac{1}{ \sum_{C' \in \cB} \mathbb{I}( (s,t) \in C' ) }
	\right\}
\end{align}
Note that $0 \leq N_e \leq N$, and if $\cB$ contains $r$ distinct bimodules with disjoint index sets (and arbitrary multiplicity) in $\cB$ then $N_{e} = r$.  With $N_e$ in hand, we apply agglomerative hierarchical clustering to 
the bimodules in $\cB$ using average linkage and a dissimilarity measure equal to the Jaccard distance
between the index sets,
\[	
d_J(C, C') = 1 - \frac{| C_1 \cap C_2 |}{ |C_1 \cup C_2 |} .\]
We then prune the resulting dendrogram by selecting the horizontal cut that yields closest to $N_e$ clusters.
Finally, from each of the resulting clusters $\cC \subseteq \cB$ we select a representative bimodule
$C \in \cC$ maximizing the centrality score
\begin{equation}
	\eta(C : \cC) = \sum_{(s,t) \in C} \,  \sum_{C' \in \cC} \mathbb{I} ((s,t) \in C').
	\label{eq:impscore}
\end{equation}
The centrality score $\eta(C : \cC)$ favors bimodules $C$ whose elements are contained in many of the other
bimodules $C'$ falling in the cluster $\cC$.

\subsection{Network-Based Assessment of Minimality}
\label{sec:connectivity}

A stable bimodule is minimal if it does not properly contain another stable bimodule.  
Minimal bimodules represent indecomposable structures, which may enhance their
interpretability in exploratory tasks.
Initializing BSP with singletons encourages the discovery of minimal bimodules, but 
in general, the bimodules output by BSP are not guaranteed to be minimal. 
As verifying minimality in practice can be computationally prohibitive, we adopt an alternative,
network based approach that assesses the decomposability of stable bimodules found by BSP. 

We begin with the cross-correlation network $G_r$, which is a weighted
bipartite network with vertex set $S \cup T$ and edge set $E = S \times T$.  Each edge
$(s,t)$ has weight $w(s,t)$ equal to the observed sample correlation $r(s,t)$ between 
features $\bbX_s$ and $\bbY_t$.  Note that $G_r$ has no edges between features of the same type.  
For each $\tau \geq 0$, let $G_r(\tau)$ be the subgraph of $G_r$ obtained by removing edges
$(s,t)$ with absolute weight $|w(s,t)| < \tau$; let $E(\tau)$ denote the resulting set of edges.
We call a pair of feature sets $(A,B)$ connected in $G_r(\tau)$
if there is a path (a sequence of adjacent edges in $E(\tau) \cap (A \times B)$) connecting every distinct pair of indices
$u,v \in A \cup B$ that goes only through vertices in $A \cup B$.

\begin{definition}
The \emph{connectivity threshold} $\tau^*(A, B)$ of a feature set pair $(A,B)$ is the largest 
$\tau \geq 0$ such that $(A,B)$ is connected in $G_r(\tau)$.  The essential edges of
$(A,B)$ are the pairs $(s,t) \in A \times B$ such that $|r(s,t)| \geq \tau^*(A, B)$.
\label{def:ess-edges}
\end{definition}

The consideration of connectivity thresholds and essential edges in our methodology is motivated by insights from large sample analysis \citep{dewaskar2021high}. This analysis reveals that in the large sample limit, minimal stable bimodules correspond to connected components of the population cross-correlation network, which consists of feature pairs $(s,t) \in S \times T$ with non-zero correlations at the population level.

The threshold $\tau^*(A,B)$ measures the strength of the weakest link required to maintain connectivity of the features $A \cup B$ in $G_r$: 
the set $A \cup B$ cannot be partitioned into two non-empty groups such that there are only weak edges
(with magnitude less than $\tau^*(A,B)$) between the groups. 
In this way $\tau^*(A,B)$ quantifies the ease with which the bimodule $(A,B)$ can be 
decomposed into a disjoint union of smaller bimodules.
In practice, the bimodules found by BSP have relatively high connectivity thresholds (e.g.~see Section \ref{sec:network-interpretation}), 
indicating that these bimodules are robustly connected.

One may also regard $E(\tau) \cap (A \times B)$ as an estimate of the feature pairs $(s,t) \in A \times B$ that are truly correlated at the population 
level, with larger values of $\tau$ providing more conservative estimates.
The value $\tau = \tau^*(A,B)$ is the most conservative threshold subject to the constraint that 
$A \cup B$ is connected in $G_r(\tau)$, and in this case the essential edges are those of the resulting graph.

\subsection{Choice of $\alpha$ Using Half-Permutation Based Edge-Error Estimates}

\label{sec:howtochoosealpha}

To select the false discovery parameter $\alpha \in (0,1)$ for BSP, we estimate the \emph{edge-error} for each value of $\alpha$ 
from a pre-specified grid. The edge-error is a network-based notion of false discovery for bimodules, defined as 
the average fraction of erroneous essential-edges (see Definition \ref{def:ess-edges} above) among bimodules. 
Since we do not know the ground truth, we estimate the edge-error for BSP by running it on instances of the 
\emph{half-permuted} dataset in which the sample labels for half of the features from each data type have been 
permuted. Further details are given below.

\subsubsection{Half-Permutation}
\label{sec:half-perm}

Running BSP on the permuted data (Definition \ref{def:perm-dist}) allows us to assess the false discoveries from BSP when there are no true associations between features from $S$ and $T$. 
However, we expect that there are true associations between features from $S$ and $T$.  
Indeed, these are the ones we wish to find. 
To create a null distribution where some pairs of features from $S$ and $T$ are correlated and some are not, we employed 
a \emph{half-permutation} scheme. Let $(\bbX, \bbY)$ denote the original data. 
We generate a \emph{half-permuted} dataset $(\tilde{\bbX}, \tilde{\bbY})$ as follows:
\begin{enumerate}
	\item Randomly select half of the features, $\hat{S} \subseteq S$ and $\hat{T} \subseteq T$, from each data type. 
	\item  Consider the matrix $\tilde{\bbX}$ obtained by randomly permuting the rows of the submatrix of $\bbX$ corresponding to features in $\hat{S}$. In other words, the sub-matrix corresponding to the features $S \setminus \hat{S}$ is the same in $\bbX$ and $\tilde{\bbX}$, while the sample labels for the submatrix of  $\tilde{\bbX}$ corresponding to features in $\hat{S}$ have been randomly permuted, i.e. $\tilde{\bbX}_{\hat{S}} = \bP_1 \bbX_{\hat{S}}$ where  $\bP_1 \in \{0,1\}^{n \times n}$ is a random permutation matrix.
	\item Similarly permute the rows of the sub-matrix of $\bbY$ corresponding to the features $\hat{T}$ using another independent permutation matrix $\bP_2 \in \{0,1\}^{n \times n}$. Call the resulting matrix $\tilde{\bbY}$.
\end{enumerate}
If any covariates are present, we correct for them after the above step. Together, the \emph{half-permutation} in steps 2 and 3 nullify the cross-correlation between pairs of features in $\hat{S} \times T \cup S \times \hat{T}$.

\subsubsection{Estimating the Edge-Error Using Half-Permutations}

Let $\cB = (A_1, B_1), (A_2, B_2), \ldots, (A_K, B_K)$ be the collection of bimodules obtained by running BSP on the half-permuted data $(\tilde{\bbX}, \tilde{\bbY})$. We define the edge-error estimate for the collection $\cB$ as
\begin{equation}
	\label{eq:edeerrest}
	\widehat{\EDGEERR}(\cB) = \frac{1}{|\cB|}\sum_{(A, B) \in \cB} \frac{\abs{\essedges(A, B) \cap \brR*{\hat{S} \times T \cup S \times \hat{T}} }}{\abs{\essedges(A, B)}},
\end{equation}
where the $\essedges(A,B)$ denotes the essential-edges (Definition \ref{def:ess-edges}) for the feature set pair $(A,B)$  based on the half-permuted data $(\tilde{\bbX}, \tilde{\bbY})$. Intuitively, the edge-error estimate captures the probability that a randomly chosen essential-edge from a randomly-chosen bimodule falls within the nullified feature pairs $\hat{S} \times T \cup S \times \hat{T}$.  Small edge-error estimates provide confidence that the bimodules detected by BSP are not being driven by spurious correlations.

We use the edge-error estimates from multiple half-permuted datasets to tune the false discovery parameter $\alpha \in (0,1)$. 
First, we generate a number of half-permuted datasets.  Then, for each $\alpha$ value in a grid of values, for example $\{0.01, 0.02, \ldots, 0.05 \}$, we run $\BSP$ with parameter $\alpha$ over each of the half-permuted datasets and calculate an average edge-error value for that $\alpha$. 
Finally, we choose an $\alpha$ from the grid that has average edge-error smaller than a pre-determined value like 0.05. 
Typically, smaller values of $\alpha$ have smaller edge-error, so we choose the largest value of $\alpha$ from the grid that has the acceptable edge error.  One may also choose smaller values of $\alpha$ to limit the sizes of bimodules. In our simulation study, this strategy to tune $\alpha$ successfully controlled the true edge-error under $0.05$.

In practice, the edge-error estimates may be quite variable even when averaged over a large number of half-permuted datasets, due
in part to the fact that different $\hat{S}$ and $\hat{T}$ are picked for each instance of the half-permutation. 
When we observed such variability, we chose a more conservative value of $\alpha$.

\subsection{Practical Workflow}
\label{sec:workflow}

Bimodules discovered by BSP can be used similarly as gene modules \citep[e.g.][]{langfelder2008wgcna} 
for downstream data exploration and hypothesis generation.
A typical workflow will start by using the R package that implements the BSP pipeline, including selection of $\alpha$, 
running BSP, post-processing bimodules, and obtaining essential-edge networks. 

One may then identify bimodules of potential relevance to the task at hand. 
For instance, in eQTL applications (discussed below), one may identify bimodules enriched for known gene 
sets (from the Gene Ontology database) that are associated with biological processes of interest.  
If additional clinical outcomes (or measurement modalities) are available for the individuals under study, 
one can score BSP bimodules based on the association between the genes in the bimodule and these outcomes.

Finally, one can apply a variety of analyses, such as identification of hub-nodes or clustering coefficients, 
to the essential-edge networks derived from the selected bimodules to further illuminate cross-correlation 
relationships that may be of interest for further study.

\section{Simulation Study}
\label{sec:sim}

To assess the performance of BSP and related methods, we simulated a dataset consisting of $n=200$ samples with $p \approx 14 \times 10^4$ and $q \approx 26 \times 10^3$ features of Types 1 and 2, respectively. (The values of $p$ and $q$ were chosen to match the number of SNPs and genes in a previous version of the GTEx thyroid dataset.)
The data was generated from a model with $K=1000$ disjoint \emph{target} bimodules of various sizes, (intra- and cross-) correlation strengths, and network structures at the population level. Previously, simulation studies incorporating fewer than ten embedded bimodules have been conducted for methods based on
sCCA \citep{waaijenborg2008quantifying, parkhomenko2009sparse, witten2009penalized} 
and graphical models \citep{cheng2016sparse, cheng2015fast}.  Here, motivated by the eQTL analysis application in 
Section \ref{sec:eqtl}, we considered a more sophisticated simulation model that incorporates a diverse collection 
of target bimodules and related network structures. To make the recovery of these bimodules more challenging, 
edges were added at random between target bimodules so that the population cross-correlation network has a 
so-called giant connected component.

\subsection{Data Model}
\label{sec:sim-data-model}
We now describe the simulation model in more detail. Following the notation at the beginning of Section \ref{sec:setup}, we denote the two types of features by index 
sets $S = \{ s_1, s_2 \ldots s_p \}$ and $T = \{t_1, t_2 \ldots t_q \}$.
For each of the $n$ individuals, the joint $p+q$ dimensional measurement vector is independently 
drawn from a multivariate normal distribution with 
mean $0 \in \R^{p+q}$ and $(p+q) \times (p+q)$ covariance matrix $\Sigma$. The covariance matrix $\Sigma$ is designed so that 
it has $K=1000$ target bimodules of various sizes, network structures, signal strengths, and intra-correlations. 

As it is difficult to generate structured covariance matrices while maintaining 
non-negative definiteness, we instead specify a generative 
model for the $p+q$ dimensional random row vector $(X, Y) \sim \mathcal{N}_{p+q}(0, \Sigma)$.
To begin, we partitioned the first-half of the $S$-indices $\{ s_1, \ldots, s_{\ceil{p/2}} \}$ into $K$ disjoint subsets 
$A_1, A_2, \ldots, A_K$ with sizes chosen according to a Dirichlet distribution with parameter $(1, 1, \ldots, 1) \in \R^K$. In the same way, we generated a Dirichlet partition $B_1, B_2, \ldots, B_K$ of the first-half of $T$ indices $\{ t_1, \ldots, t_{\ceil{q/2}} \}$ independent of the previous partition.  The feature set pairs $(A_i, B_i)$ constitute the target bimodules, while the features in second-half of the $S$- and $T$-indices are not part of any bimodules. 
Next, the random sub-vectors $(X_{A_i}, Y_{B_i})$ corresponding to the target bimodules were 
generated independently for each $i \in [K]$ using a graph based regression model described next.

Let $(A,B)$ be a feature set pair, and suppose that $\rho \in [0,1)$ and $\sigma^2 > 0$ are given. Let $D \in \{0,1\}^{|A| \times |B|}$ be 
a binary matrix, which we regard as the adjacency matrix of a connected bipartite network with vertex set $A \cup B$.  Then the random row-vector $(X_{A}, Y_{B})$ is
generated as follows:
\begin{equation}
	\label{eq:reg}
	X_A \sim \mathcal{N}_{\abs{A}}(0, (1-\rho) I + \rho U) \ \mbox{ and } \ Y_{B} =  X_A  D + \epsilon,
\end{equation}
where $\epsilon \sim \mathcal{N}_{\abs{B}}( 0, \sigma^2 I)$ and $U$ is a matrix of all ones. 
To understand the bimodule signal produced by this model, note that $\rho$ governs the intra-correlation between features in $A$ and that
for any $t \in B$, the variable $Y_{t}$ is influenced by features $X_{s}$ such that $(s,t)$ is an edge in the
adjacency matrix $D$. 
For each of the target bimodules $(A_i, B_i)$ in the simulation, we independently chose parameters $\rho_i$, $\sigma_i^2$, and $D_i$ to produce a variety of behaviors while maintaining the inherent constraints between them. See Section \ref{sec:sim-model-details} for more details.

Features $X_{s_j}$ with $j >  \ceil{p/2}$ are independent $\mathcal{N}(0,1)$ noise variables.
Features $Y_{t_r}$ with $r >  \ceil{q/2}$ are either noise (standard normal) or they are bridge variables that connect two target bimodules.  
In more detail, for every pair of distinct bimodules $(A_k, B_k)$ and $(A_l, B_l)$ with 
$1 \leq k < l \leq K$, with probability
 $\kappa=\frac{1.5}{K}$, 
we connect the two bimodules by selecting 
at random (and without replacement) an index $r >  \ceil{q/2}$ and making it a bridge variable by defining 
\begin{equation}
	Y_{t_r} = X_{s} + X_{s'} + \epsilon \quad\text{ with } \epsilon \sim N(0, \sigma_r^2)
	\label{eq:bridge},
\end{equation} 
for a randomly chosen $s \in A_k$ and $s' \in A_l$. Notice that the bridge variable $Y_{t_r}$ now plays the role of connecting the target bimodules $(A_k, B_k)$ and $(A_l, B_l)$ in the \emph{population cross-correlation network}, given by vertex set $S \cup T$ and edges $(s,t) \in S \times T$ such that there is non-zero (population) correlation between features $X_s$ and $Y_t$. The noise variance $\sigma_r^2$ in \eqref{eq:bridge} is chosen so that the correlation between 
$Y_{t_r}$ and $X_s$ (and $X_{s'}$) 
is equal to the average of the maximum cross-correlation of the bimodules that are being connected.  
Finally, if $Y_{t_r}$ is not a bridge variable, it is taken to be noise (standard normal).

Prior to the addition of bridge variables, the connected components of the population cross-correlation network are just the bimodules
$(A_i, B_i)$ for $i = 1, \ldots, K$.  Once bridge variables have been added, the population cross-correlation network will have a 
so-called giant connected component
comprising a substantial portion of the underlying index space $S \cup T$.  
While theoretical support for the presence of giant component in our simulation model 
comes from the study of Erd\"os-Renyi random graphs \citep{bollobas2001evolution},  
such components have also been observed in empirical eQTL networks 
\citep{fagny2017exploring, platig2016bipartite}.   Since we only add a relatively small number (752) of bridge variables,  the majority of the cross-correlation signal is still contained in the more densely connected sets $(A_i, B_i)$, $i=1, \ldots, K$.

\subsubsection{Simulation Model for Each Target Bimodule}
\label{sec:sim-model-details}

Here we describe further details on how the parameters $\rho, \sigma \in [0,1]$ and $D$ that appear 
in \eqref{eq:reg} are chosen for each target bimodule $(A,B)$.  
First, we introduce two new parameters $\beta, \eta \in [0,1]$, where $\beta$ is the density of edges in the graph $D$, and $\eta$ denotes the cross-correlation between features from $B$ and adjacent features from $A$ in the graph $D$. Together, the parameters $\beta, \rho, \eta \in [0,1]$ respectively control the network connectivity, intra-, and cross-correlation strengths of the 
target bimodule $(A, B)$, and will be used to generate appropriate $D$ and $\sigma^2$. For each target bimodule $(A, B)$, these values are independently sampled using the following steps:

\begin{enumerate}
	\item Choose a constant $\beta \in [0,1]$ uniformly at random. With $d \doteq \ceil{\beta \abs{A}}$, let $D$ be the adjacency matrix of a $d$-regular bipartite graph on vertex sets $A$ and $B$ formed by independently connecting each vertex $t \in B$  to $d$ randomly chosen vertices from $A$. We want to ensure that the graph $D$ is connected. If it is not connected, we repeatedly increase $\beta$ to $\beta + \Delta \beta$ where $\Delta\beta = 0.1$ and re-instantiate the graph $D$ until it is connected.
	\item Randomly choose $\rho \in [0,1]$ and $\eta \in [0,.8]$ subject to the constraint $\delta \doteq 1 + \rho(d-1) \geq \eta^2 d$. We satisfy this constraint by first uniformly generating $\rho$ and then generating $\eta$ uniformly from $[0, \min(\sqrt{\delta d^{-1}},.8)]$.
	\item Finally let $\sigma = \frac{\sqrt{\delta(\delta - \eta^2d)}}{\eta}$.
\end{enumerate} 
Lemma \ref{lem:regressor} in \Cref{sec:sim-appendix} shows that, with this choice of parameters, features connected 
by the adjacency matrix $D$ have population cross-correlation equal to $\eta$.

\subsection{Evaluation and Comparison of Methods}

We ran BSP, CONDOR, sCCA, and \meqtl\ on the simulated dataset. 
The parameter $\alpha=0.02$ for BSP was chosen to keep the edge-error estimates 
based on half-permuted data (see \Cref{sec:howtochoosealpha}) under $0.05$. 
The q-value cutoff for \meqtl\ was also taken to be $0.05$.
More details on how the various methods were run are provided in \smref{sec:sim-appendix}.

\subsubsection{Assessment Metrics}
We compared the results from each method to the ground truth to assess (a) the recovery of target bimodules by each method, 
and (b) the presence of spurious associations within the bimodules detected by these methods.  We considered the following metrics.
\begin{enumerate}
	\item \emph{Jaccard similarity for each target bimodule}. We calculated the largest Jaccard similarity between each target bimodule and a detected bimodule.  Jaccard similarity is computed by regarding a bimodule $(A,B)$ as the set of pairs $A \times B$.
	\item \emph{Recall for each target bimodule.} We calculated the largest fraction of pairs from each target bimodule that were included in a detected bimodule. 
	\item \emph{Edge-error for each detected bimodule.} The \emph{edge-error} of a detected bimodule is defined as the fraction of essential-edges (Definition \ref{def:ess-edges}) from the detected bimodule that are false, i.e.~not contained within any true bimodule or the set of confounding edges. \end{enumerate}
The Jaccard and recall metrics above provide two different ways to assess recovery of target bimodules, 
while edge-error provides a way to assess spurious associations. 
While the Jaccard similarity measures the exact recovery of target bimodules, 
recall measures recovery relative to inclusion.  In defining edge-error, we focus on false discoveries 
among essential-edges, since all pairs within a bimodule need not be correlated.

Since \meqtl\ only finds significant feature pairs, and not bimodules, for this method 
recall is defined as the fraction of pairs from each target bimodule that are 
discovered by the method, while edge-error is defined as the fraction of pairs detected by the method that are false.

\subsubsection{Results From Evaluation}

\begin{figure}
	\begin{subfigure}{0.5\textwidth}
		\centering    
		\includegraphics[width=\textwidth]{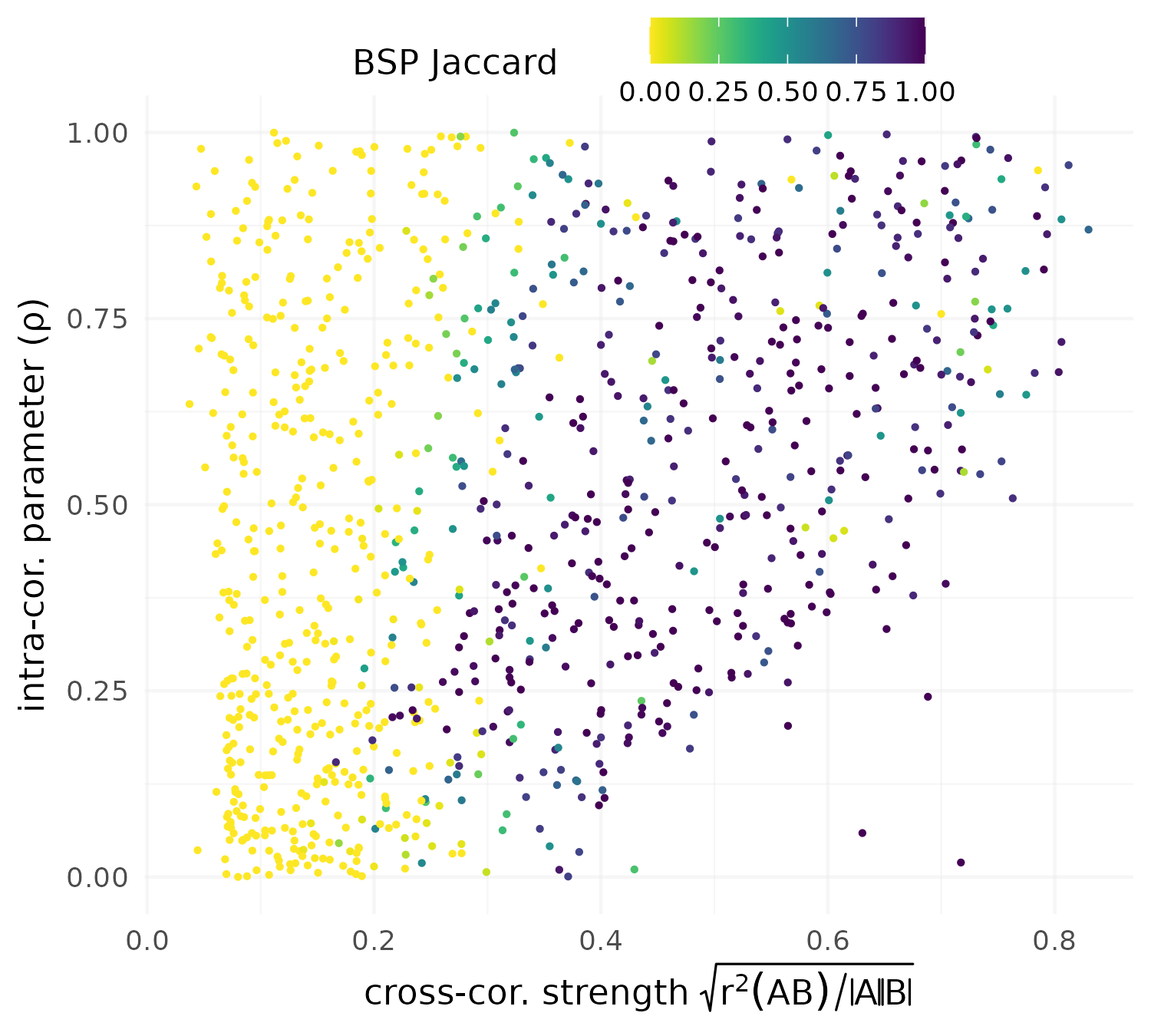}
	\end{subfigure}
	\begin{subfigure}{0.5\textwidth}
		\centering 
		\includegraphics[width=.9\textwidth]{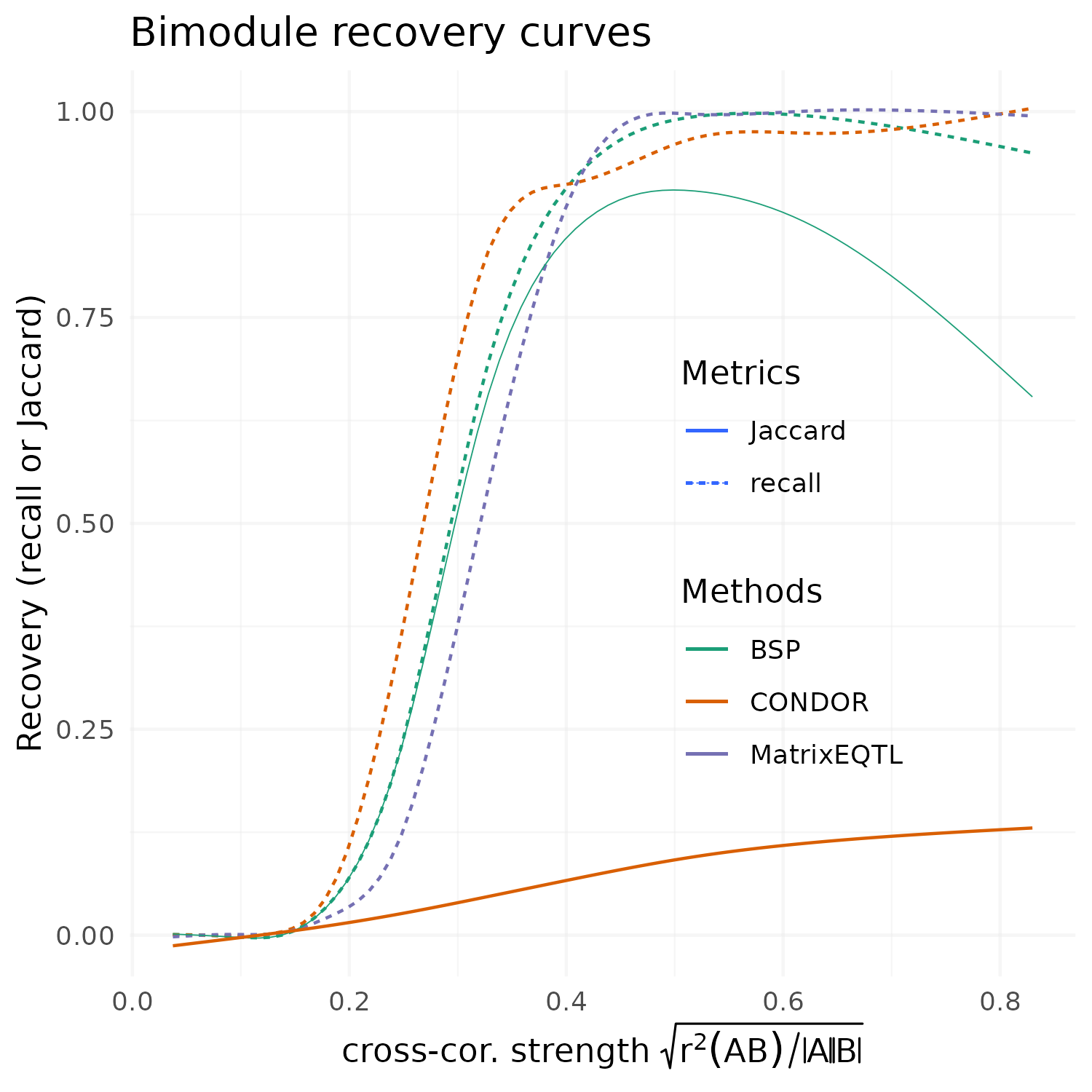}   
	\end{subfigure}
	\caption{Recovery of target bimodules under the equality based metric Jaccard and the inclusion based metric recall. Left: dependence of cross-correlation strength and intra-correlation parameter of target bimodules on BSP Jaccard. Right: the averaged recovery curves for target bimodules under CONDOR, BSP, and \meqtl\ as a function of their cross-correlation strength.}
	\label{fig:recall}
\end{figure}

Figure \ref{fig:recall} (left) shows the recovery of target bimodules by BSP.  The performance of BSP was 
influenced primarily by the cross-correlation strength $\sqrt{\frac{r^2(A,B)}{|A||B|}}$ of the target bimodule, 
though the intra-correlation strength, measured by the parameter $\rho$ appearing in \eqref{eq:reg}, 
also had an effect. 	
Most target bimodules with cross-correlation strength above 0.4 were completely recovered, 
while those with strength below 0.2 were not recovered. 
For strengths between 0.2 to 0.4, there was a variation in recovery, with smaller Jaccard recovery for 
bimodules having larger intra-correlation strength.  The effect of intra-correlation strength on recovery was expected 
as BSP accounts for the intra-correlations among features of the same type. 
BSP does a good job of controlling false discoveries: 
the average edge-error for BSP bimodules was 0.041, and 90\% of the BSP-bimodules had edge-error under 0.11.

The results from sCCA, CONDOR, and \meqtl\ on the simulation study are described in detail in \smref{sec:sim-appendix}, but we summarize the important results here.  
The recovery of target bimodules by CONDOR was rather insensitive to 
intra-correlation strength, so we only considered the effect of cross-correlations.  As CONDOR bimodules often contained 
multiple target bimodules, it had better performance in terms of recall than Jaccard similarity. For the recall metric, the average recovery curves for CONDOR, \meqtl, and BSP, as a function of the cross-correlation strength, were comparable (Figure \ref{fig:recall}, right). 
CONDOR bimodules had an average edge-error of $0.09$.
Finally, \meqtl\ found 436,616 significant pairs, 10\% of which were false discoveries. The inflated type-1 error for the last two methods (despite using the q-value cutoff of $0.05$) may be due to the fact that these methods do not account for intra-correlations.

Concerning sCCA, the detected bimodules were very large and had a high average edge-error of $0.94$
when we used the default parameter choice.  We also ran the procedure with a range of parameters that yielded smaller bimodules, but
the recovery and edge-error of the procedure continued to lag behind those of BSP and CONDOR.

We also studied the performance of BSP and CONDOR on a simulation study with larger sample size $n=600$ (see \smref{sec:large-sample-sim}). 
As expected, both methods were able to recover bimodules with lower cross-correlation strengths than earlier, in terms of recall. However, in terms of Jaccard similarity, both BSP and CONDOR had lower recovery than in the $n = 200$ simulation. We briefly discuss this behavior in Section \ref{sec:discussion}.  

\section{Application of BSP to eQTL Analysis}
\label{sec:eqtl}

Much of the existing work on bimodules is focused on the integrated analysis of genomic data.
In this section we present an extended application of BSP to expression quantitative trait loci (eQTL) 
analysis based on data from the Genotype Tissue Expression (GTEx) project. An application of BSP to discover regions of temperature and precipitation correlations in North America can be found in Appendix \ref{sec:climate}.

The next subsection provides a brief overview of eQTL analysis. The application of BSP to this problem is discussed in 
\Cref{sec:gtex-application}.

\subsection{Expression Quantitative Trait Loci Analysis}
\label{sec:eqtl-intro}

Genetic variation within a population is commonly studied through single nucleotide 
polymorphisms, referred to as SNPs.  A SNP is a particular location in the genome where there 
is at least moderate variation in the paired nucleotides among members of a population. 
The value of a SNP for an individual is the number of reference nucleotides appearing at that site, 
which takes the values 0, 1, or 2.  After normalization and covariate 
correction, the value of a SNP may no longer be discrete.

eQTL analysis seeks to identify SNPs that affect the expression of one or more genes.
A SNP-gene pair for which the expression of the gene is correlated with the value of the SNP is referred to as an eQTL. 
Identification of eQTLs is an important first step in the study of genomic
pathways and networks that underlie disease and development 
in human and other populations \citep[see][]{nica2013expression,albert2015role}. 

In modern eQTL studies it is common to have measurements of 10-20 thousand genes and 2-5 million SNPs on 
hundreds (or in some cases thousands) of samples.  
Identification of putative eQTLs or genomic ``hot spots'' is carried out by evaluating the correlation of 
numerous SNP-gene pairs, and identifying those meeting an appropriate multiple testing based threshold.  
In studies with larger sample sizes 
it may be feasible to carry out \transeqtl{} analyses, which consider all SNP-gene pairs regardless of genomic location.  
However, it is more common to carry out \ciseqtl{} analyses, in which one restricts attention to SNP-gene pairs for which the SNP 
is within some fixed genomic distance (often 1 million base pairs) of the gene's transcription start site, and in particular, on the same chromosome \citep[c.f.][]{ westra2014genome, gtex2017genetic}. We use the prefixes \textit{cis}- and \textit{trans}- to refer to the type of eQTL analyses, while using adjectives \emph{local} and \emph{distal} to denote the proximity of the discovered SNP-gene pairs. In particular, \ciseqtl{} analyses seek to discover local eQTLs, while \transeqtl{} analyses seek to discover 
\emph{both} local and distal eQTLs.

As a result of multiple testing correction needed to address the large number of SNP-gene pairs under study, 
both \textit{trans}- and \textit{cis}-eQTL analyses can suffer from low power.
Several methods have been proposed to improve the power of standard eQTL analyses, including penalized 
regression schemes that try to account for intra-gene or intra-SNP interactions \citep[and references therein]{tian2014methods}, 
and methods that consider gene modules as high-level phenotypes to reduce the 
burden of multiple-testing \citep{kolberg2020}.
For instance, \cite{huang2009graph} proposed a network-based approach for improving the discovery of eQTLs by creating a tripartite graph involving genes, SNPs, and samples. They utilized maximal cliques as a heuristic to reduce the search space over SNP-gene pairs to test for eQTL associations.

As an alternative, one may shift attention from individual SNP-gene pairs to SNP-gene bimodules.
\cite{cheng2015fast,cheng2016sparse} refer to such bimodules as ``group-wise eQTLs''.
As genes often act in concert with one another, bimodule discovery methods 
can gain statistical power from group-wise interactions, by borrowing strength across 
individual SNP-gene pairs. Further, it is known that activity in a cell may be the result of a regulatory network of genes rather than individual genes \citep{chakravarti2016revealing}. Hence, bimodules may represent a group of SNPs that disrupt the functioning of gene regulatory networks and contribute to diseases \citep{platig2016bipartite}.

\subsection{eQTL Analysis of GTEx Thyroid Data}
\label{sec:gtex-application}

Here we describe the application of bimodules to the problem of expression quantitative trait loci (eQTL) analysis. 
The NIH funded GTEx Project has collected and created a large eQTL database containing 
genotype and expression data from postmortem tissues of human donors.  
We applied BSP, CONDOR, and standard eQTL-analysis (using \meqtl) to $p = 556,304$ SNPs and 
$q = 26,054$ thyroid expression measurements from $n=574$ individuals.  
A detailed account of data acquisition, preprocessing, and covariate correction, along with additional details about the results and analysis in this section, can be found in Appendix \ref{sec:gtex-appendix}.  

\subsection{Running BSP and Other Methods}

We applied BSP to the thyroid eQTL data with false discovery parameter $\alpha = 0.03$, selected using a permutation-based procedure 
to keep the edge-error estimates under $0.05$. 
The search \Cref{alg:bsp} was run $p/2 + q \approx$ 300K times starting from initial conditions consisting 
of all genes and half of the randomly chosen SNPs. 
The majority ($\approx$277K) of these searches found an empty fixed point within the first few iterations. 
Of the remaining 27K searches, the great majority identified a non-empty fixed point within 20 steps. 
Only 20 searches cycled and did not terminate in a fixed point. 
The search produced 3744 unique stable bimodules; the effective number of bimodules was 3304.  
We applied the filtering procedure described in Section \ref{sec:post-processing} to select a subfamily of 3304 bimodules that were substantially disjoint.  

We also performed standard \emph{cis-} and \emph{trans-}eQTL analysis on the thyroid eQTL data using \meqtl\ \citep{matrixEQTL}, and applied CONDOR \citep{platig2016bipartite} and sCCA \citep{witten2009penalized} to produce bimodules. 
CONDOR produced six bimodules in total, while sCCA was tasked with identifying 50 bimodules. 
Figure \ref{fig:size} shows the sizes of the bimodules identified by the various methods. 
All bimodules identified by sCCA were very large, making them difficult to analyze and interpret.  
The identified bimodules also exhibited moderate overlap (the effective 
number was 25).  As such, we excluded the sCCA bimodules from subsequent comparisons. 
Analysis of sCCA on the simulated data suggests that the method may be able to 
recover smaller bimodules with a more tailored choice of its parameters, but we did not pursue this here.

\subsection{Quantitative Validation}

In this subsection, we apply several objective measures to validate and understand the bimodules found by BSP and CONDOR.

\subsubsection{Permuted Data}
\label{sec:perm}

In order to assess the propensity of each method to detect spurious bimodules, we applied BSP and CONDOR to five data sets 
obtained by jointly permuting the sample labels for the expression measurements and most covariates (all except the five genotype PCs), 
while keeping the labels for genotype measurements and genotype covariates unchanged. Each data set obtained in this way is a realization
of the permutation null from Definition \ref{def:perm-dist}.
BSP found very few (5-12) bimodules in the permuted datasets compared to the real data (3344). 
CONDOR found no bimodules in any of the permuted datasets.

\subsubsection{Bimodule Sizes}

\begin{figure}[t]
	\begin{minipage}[b]{0.45\linewidth}
	\centering
	 \includegraphics[width=\linewidth]{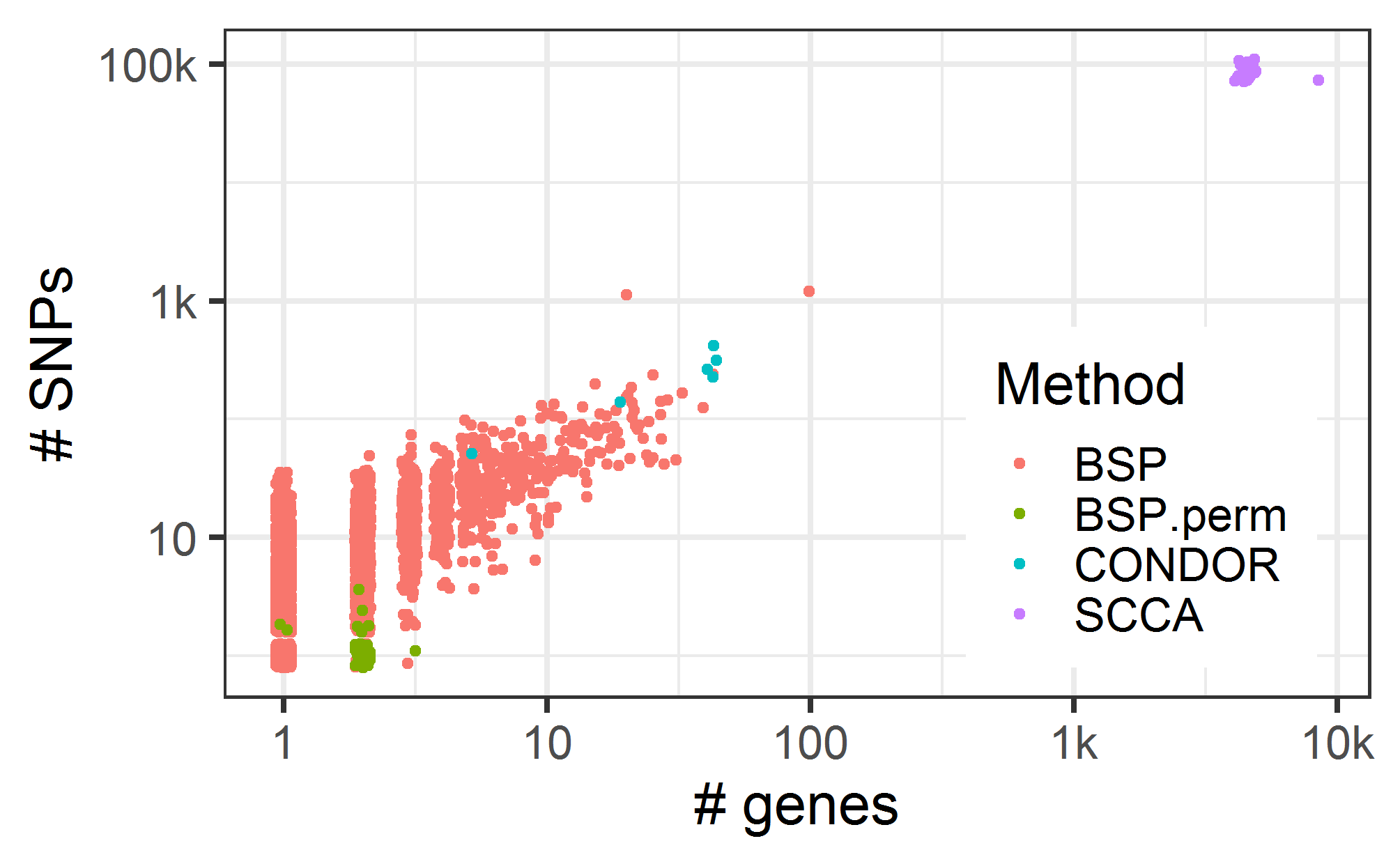}
	 \caption{\footnotesize{} The sizes of bimodules detected by BSP, CONDOR and sCCA, and 
	 sizes of bimodules detected by BSP under the 5 permuted datasets.}
	\label{fig:size}
	\end{minipage}
\hspace{0.5cm}
	\begin{minipage}[b]{0.45\linewidth}
	\centering
	  \includegraphics[width=\linewidth]{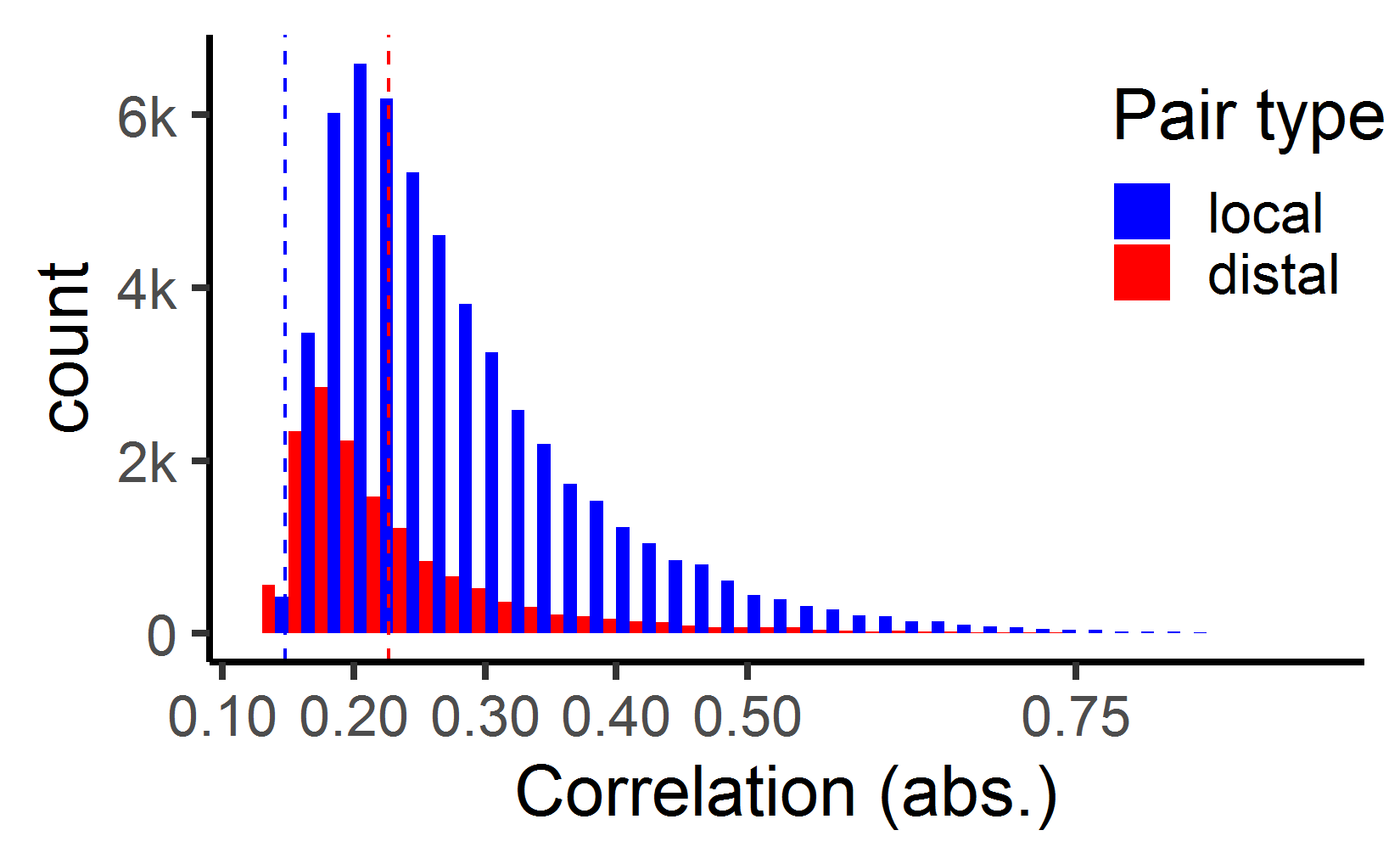}
	  \caption{\footnotesize{} BSP is able to detect weak signals. Correlations corresponding to SNP-gene pairs that appear as 
	  	essential edges (Section \ref{sec:connectivity}) in one or more BSP bimodules with mean size ($\sqrt{|A||B|}$) above 10. Local pairs to the left of the blue line (\textit{cis}-analysis threshold) and distal pairs to the left 
	  	 red line (\textit{trans}-analysis threshold) show importance at the network level but were not discovered by standard eQTL analyses.}
\label{fig:histimppairs}
	\end{minipage}
\end{figure}

Most (89\%) of the bimodules found by BSP have fewer than 4 genes and 50 SNPs, but BSP also identified moderately 
sized bimodules having 10-100 genes and 30-1000 SNPs (see Figure \ref{fig:size}). 
The bimodules found by CONDOR were moderately sized, with 10-100 genes and several hundred SNPs, 
except for one smaller bimodule with 5 genes and 43 SNPs. 
On the permuted data, most bimodules found by BSP have fewer than 2 genes and 2 SNPs.

\subsubsection{Connectivity Threshold and Network Sparsity}
\label{sec:network-interpretation}  

Stable bimodules capture aggregate association between groups of SNPs and genes.
In some cases one may wish to identify individual SNP-gene associations of interest 
within discovered bimodules. 
A natural starting point for this is the network of essential edges of the bimodule, defined in 
Section \ref{sec:connectivity}.
To better understand the structure of this network of essential edges, we calculate
the \emph{tree-multiplicity} 
\begin{equation}
  \TreeMul(A, B) \doteq \frac{|\essedges(A,B)|}{|A| + |B| - 1},
\end{equation}
which measures the number of essential edges relative to the number of edges in a tree on the same vertex set. 
$\TreeMul(A,B)$ is never less than 1, and takes the value 1 exactly when the essential edges form a tree.

For bimodules found by BSP, the connectivity thresholds ranged from 0.14 to 0.59, and 
tree-multiplicities ranged from 1 to 10
(see Figure \ref{fig:network} in Appendix \ref{sec:gtex-appendix}). 
Smaller bimodules had larger connectivity thresholds and smaller tree multiplicities.  As such, 
these bimodules had tree-like essential edge networks comprised of strong 
(and typically local) 
SNP-gene associations. 
Larger bimodules had lower connectivity thresholds and larger tree multiplicities.  As such, these
bimodules had more redundant essential edge networks comprised of weaker 
(and often distal) 
SNP-gene associations. 
While the essential edge networks for larger bimodules had tree-multiplicities around 
10,
these networks were still sparsely connected compared 
to the complete bipartite graph on the same nodes.

\subsection{Biological Validation}
 
In order to assess the potential biological utility of bimodules found by BSP, we compared the SNP-gene pairs 
in bimodules to those found by standard \textit{cis-} and \textit{trans-}eQTL analyses.  In addition, we 
studied the locations of the SNPs, and examined the gene sets for enrichment of known functional categories.

\subsubsection{Comparison With Standard eQTL Analysis}

As described earlier, the bimodules produced by CONDOR are derived in a direct way from SNP-gene pairs
identified by \textit{cis}- and \textit{trans}-eQTL analyses.     
Table \ref{tab:standard-eQTL-comp} compares the eQTL pairs identified by standard analyses 
with those found in bimodules identified by BSP.  
Recall that \ciseqtl{} analysis considers only local SNP-gene pairs, which improves detection power by
reducing multiple testing,
while \transeqtl{} analysis and BSP do not use any information about the absolute or relative genomic
locations of the SNPs and genes.
We find that half of the pairs identified by \ciseqtl{} analysis and most of the pairs identified by \transeqtl{}
analysis appear in at least one bimodule.

Bimodules capture subnetworks of SNP-gene associations rather than individual eQTLs, 
and as such all SNP-gene pairs in a bimodule need not be eQTLs. In fact, as noted above, the 
detected networks underlying large bimodules are typically sparse relative to the complete
network on the genes and SNPs of the bimodule.
A bimodule $(A,B)$ is connected by a set of eQTLs if the bipartite graph with 
vertex set $A \cup B$ and edges restricted to the set of eQTLs is connected.
As shown in Table \ref{tab:standard-eQTL-comp}, a substantial fraction of BSP 
bimodules are not connected by SNP-gene pairs obtained by \ciseqtl{} or \transeqtl{} analyses. 
The discovery of such bimodules suggests that the subnetworks identified by BSP cannot be found by simple  
post-processing of results from standard eQTL analyses.  
Hence, the subnetworks identified by BSP may provide new insights and hypotheses for further study.

To identify potentially new eQTLs using BSP, we examine bimodule connectivity under the combined 
set of \textit{cis-} and \textit{trans-}eQTLs.  All of the bimodules with one SNP or one gene are connected by the
combined set of eQTLs,
and therefore all edges 
in these singleton bimodules are discovered by standard analyses.  On the other hand, 224 out of the 358 
bimodules with mean size ($\sqrt{|A||B|}$) larger than 10 were not connected by the combined set of eQTLs. 
In Figure \ref{fig:histimppairs}, we plot the correlations corresponding to SNP-gene pairs that appear as 
essential edges in one or more bimodules with 
mean size above 10, along with the correlation thresholds for \textit{cis}-eQTL (blue line) and 
\textit{trans}-eQTL (red line) analyses. Around 300 local edges (i.e.\ the SNP is located within 1MB of the gene transcription start site) and 8.8K distal edges do not meet the correlation thresholds for \textit{cis}- and \textit{trans}-eQTL analysis, respectively, 
but show evidence of importance at the network level, and may be worthy of further study.

    \begin{table}
        \begin{tabular}{|c|c|c|}
        \hline
        Analysis type & \% eQTLs found among bimodules & \% bimodules connected by eQTLs\\
        \hline
        \transeqtl{} analysis           &  84\%                   & 70\%                   \\
        \ciseqtl{} analysis            &  51\%                   & 88\%                    \\
        \hline
        \end{tabular}
        \caption{Comparison of BSP and standard eQTL analyses. A gene-SNP pair is said to be found among 
        a collection bimodules if the gene and SNP are both part of some common bimodule. 
        On the other hand, we say that a bimodule is connected by a collection of eQTLs 
        if the bimodule forms a connected graph when the gene-SNP pairs from the collection are treated as edges. }
        \label{tab:standard-eQTL-comp} 
    \end{table}

\subsubsection{Genomic Locations}
\label{sec:genloc}

We studied the chromosomal location and proximity of SNPs and genes from bimodules
 found by BSP and CONDOR.   
While CONDOR uses genomic locations as part of the \ciseqtl{} analysis in its first stage, 
BSP does not make use of location information. 
Genetic control of expression is often enriched in a region local to the gene \citep{gtex2017genetic}.
All CONDOR bimodules, and almost all (99.3\%) BSP bimodules, have at least one local SNP-gene  
pair, i.e.~the SNP is located within 1MB of the gene transcription start site. 
In 94\% of smaller BSP bimodules ($\sqrt{|A||B|} \leq 10$) 
and 55\% of medium to large BSP bimodules ($\sqrt{|A||B|} > 10$) each gene and each SNP had a local 
counterpart SNP or gene within the bimodule.

For each bimodule, we examined the chromosomal locations of its SNPs and genes. 
All SNPs and many of the genes from the six CONDOR bimodules were located on Chromosome 6;
two CONDOR bimodules also had genes located on Chromosome 8 and Chromosome 9. 
The SNPs and genes from the BSP bimodules were distributed across all 23 chromosomes: 
170 of the 2947 small bimodules spanned 2 to 5 chromosomes and 
152 of the 358 medium to large bimodules spanned 2 to 11 chromosomes; the remaining bimodules were localized 
to a single chromosome, which varied from bimodule to bimodule. 

\begin{figure}
\centering
\begin{subfigure}{.5\textwidth}
  \centering
 \includegraphics[width=\linewidth]{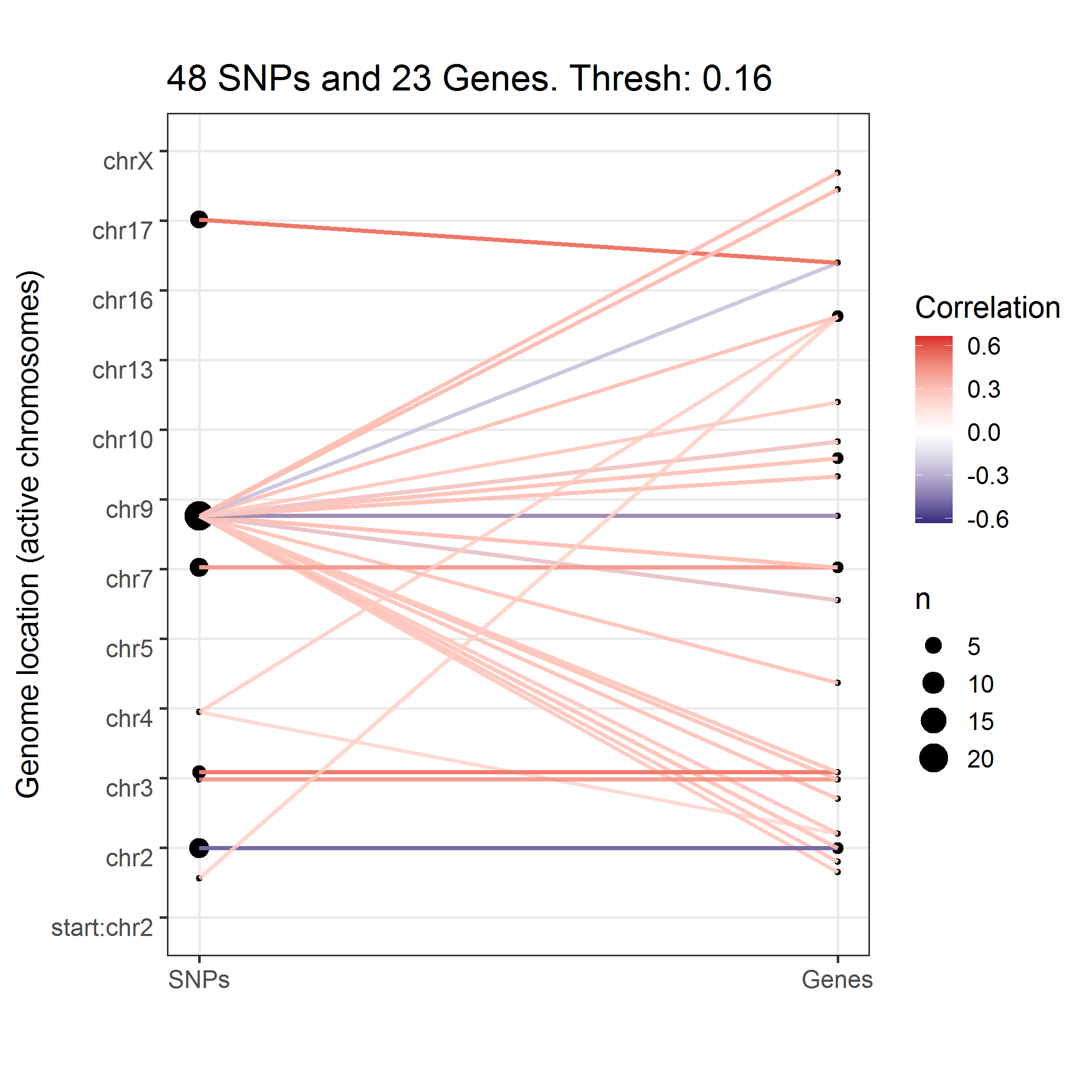}
\end{subfigure}\begin{subfigure}{.5\textwidth}
  \centering
  \includegraphics[width=\linewidth]{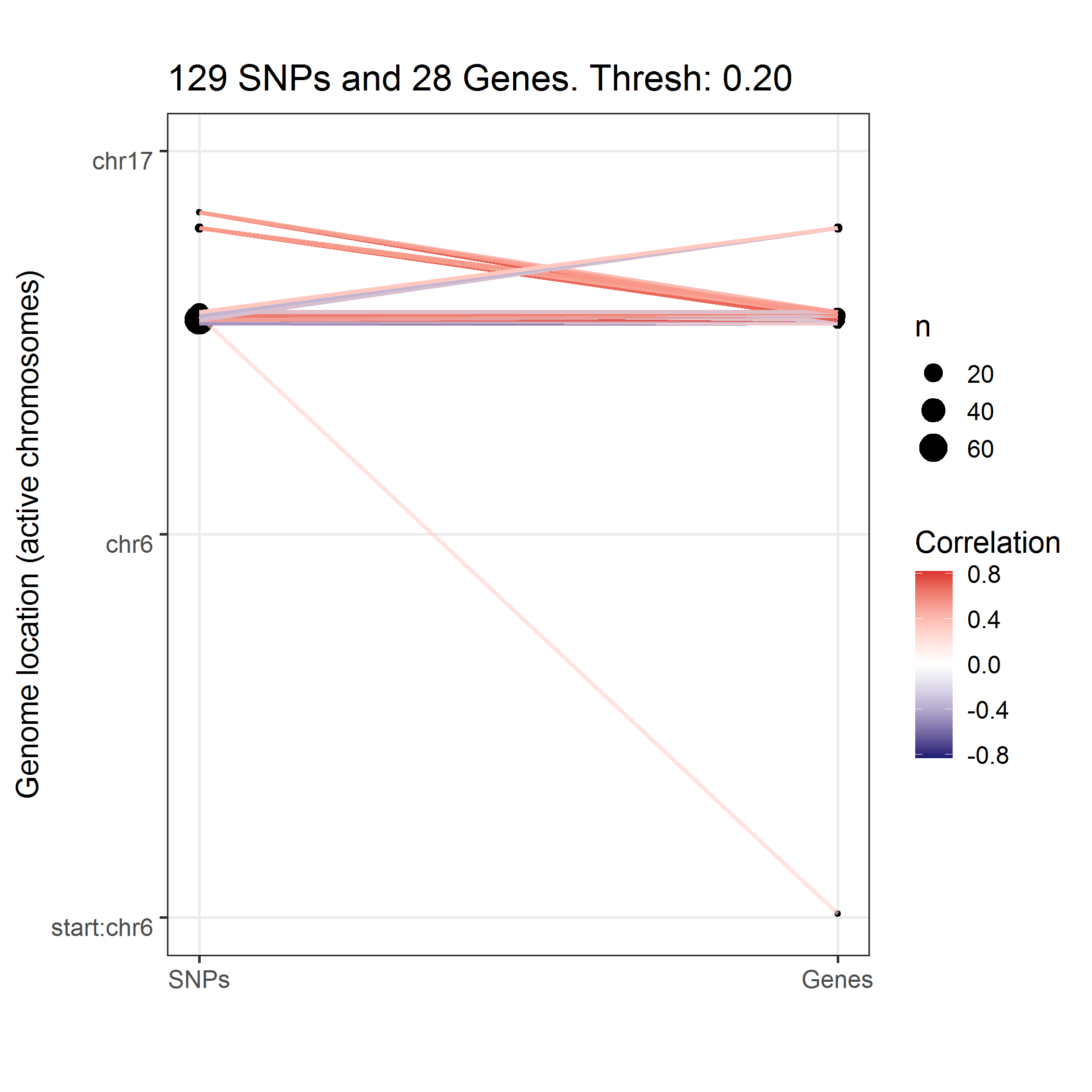}
\end{subfigure}
\caption{Two examples of SNP-gene essential edge networks from bimodules discovered by BSP, mapped onto the genome. The edges in these networks were obtained by thresholding the cross-correlation matrix for the bimodule at the connectivity threshold (Section \ref{sec:connectivity}). Comparing such networks to known gene regulatory networks may aid in identifying new SNP-gene interactions.}
\label{fig:example-bimods}
\end{figure}

Figure \ref{fig:example-bimods} illustrates the genomic locations of two bimodules found by BSP, with
SNP location on the left and gene location on the right (only active chromosomes are shown).
In addition, the figure illustrates the essential edges (Section \ref{sec:connectivity}) of each bimodule.  
The resulting bipartite graph provides insight into the underlying associations between SNPs and genes that
constitute the bimodule.  Additional examples can be found in Appendix \ref{sec:gtex-appendix}.

\subsubsection{Gene Ontology Enrichment for Bimodules} 
    
The Gene Ontology (GO) database contains a curated collection of gene sets that are known to be 
associated with different 
biological functions \citep{go2014, botstein2000gene, rhee2008use}.  
We used the topGO \citep{topGo} package to assess, via multiple Fisher's tests, which sets in the GO database are enriched within a given gene set when compared to the set of all expressed genes. For each of the 145 BSP bimodules having a gene set $B$ with 8 or more elements, we used topGO to assess the enrichment of $B$ in 
6463 GO gene sets of size more than 10, representing biological processes. 
We retained results with significant BH $q$-values ($\alpha = .05$). 
Of the 145 gene sets considered, 18 had significant overlap with one or more biological processes. 
Repeating this with randomly chosen gene sets of the same size yielded no results. 
The table of significant GO terms for BSP and CONDOR is summarized in Appendix \ref{sec:gtex-appendix}.

\section{Discussion}
\label{sec:discussion}

The Bimodule Search Procedure (BSP) is an exploratory tool that searches for pairs of feature sets with significant 
aggregate cross-correlation, which we refer to as bimodules.
Rather than relying on an underlying generative model, BSP makes use of iterative hypothesis-testing to
identify bimodules satisfying a natural stability condition. 
The false discovery threshold $\alpha \in (0,1)$ is the only free parameter of the procedure. 
Efficient approximation of the p-values used for iterative testing allow BSP to run on large datasets.

Using a complex, network-based simulation study, we found that BSP was able to recover most target bimodules 
with significant cross-correlation strength, while simultaneously controlling the false discovery of edges having network-level importance. 
Among target bimodules with moderate cross-correlation strength, BSP required the bimodules with higher intra-correlations to demonstrate higher cross-correlation strength in order to be recovered.

When applied to eQTL data, BSP bimodules identified both local and distal effects, capturing half of the eQTLs found by standard 
\textit{cis}-analysis and most of the eQTLs found by standard \textit{trans}-analysis. 
Further, a substantial proportion of bimodules contained SNP-gene pairs that were important at the network level within the BSP bimodules but 
not deemed significant under the standard (pairwise) eQTL-analyses.

At root, the discovery of bimodules by BSP and CONDOR is driven by the presence or absence of correlations between
features of different types.  A key issue for these, and related, methods is how they behave with increasing sample size. 
In general, increasing sample size will yield greater power to detect cross-correlations, and therefore one expects 
the sizes of bimodule to increase.  While this is often a desirable outcome, in applications where non-zero cross-correlations  
(possibly of small size) are the norm, this increased power may yield very large
bimodules with little interpretive value.  Evidence of this phenomenon is found in the simulation study where,
due to the presence of confounding edges between target bimodules, increasing the sample
size from $n=200$ to $n=600$ yields larger BSP bimodules, which often contain multiple target bimodules 
(\smref{sec:large-sample-sim}).  This may well reflect the underlying biology of genetic regulation:
the omnigenic hypothesis of \cite{boyle2017expanded} suggests that a substantial portion of the gene-SNP cross-correlation 
network might be connected at the population level.

An obvious way to address ``super connectivity'' of the cross-correlation network is to change the definition 
of bimodule to account for the magnitude of cross-correlations, rather than their mere presence or absence.  
Incorporating a more stringent definition of connectivity in BSP would require modifying the permutation null 
distribution and addressing the theory and computation behind such a modification, both of which are areas of
future research.

\bigskip
\begin{center}
	{\large\bf SUPPLEMENTARY MATERIAL}
\end{center}

\begin{description}
	
	\item[Appendix:]  The appendix section below contains further details on BSP implementation (Section \ref{sec:bsp-appendix}), the simulation study (Section \ref{sec:sim-appendix}), and eQTL analysis (Section \ref{sec:gtex-appendix}). We also provide a climate science application of BSP for discovering temperature and precipitation correlations in North America (Section \ref{sec:climate}).
	
	\item[BSP R package:] \url{https://github.com/miheerdew/cbce}
\end{description}

\section*{Acknowledgements}
M.D., J.P., and A.B.N.\ were supported by NIH grant R01 HG009125-01 and NSF grants DMS-1613072 and DMS-2113676. A.B.N.\ was also supported by NSF Grant DMS 2113676, and M.D. by the grant N00014-21-1-2510 from the Office of Naval Research.  
M.H.\ was awarded the Department of Defense, Air Force Office of Scientific Research, National Defense Science and Engineering Graduate (NDSEG) Fellowship, 32 CFR 168a and funded by government support under contract FA9550-11-C-0028. 
M.I.L.\ was supported by NIH grants R01 HG009125-01 and R01-HG009937. 
The authors wish to acknowledge numerous helpful conversations with Professors Fred Wright and Richard Smith. Finally, the authors wish to thank the anonymous reviewers for their careful reading of the manuscript and providing detailed suggestions that have improved our presentation.

\appendix


\section{BSP implementation details}
\label{sec:bsp-appendix}

\subsection{Dealing With Cycles and Large Sets}
\label{sec:extract-loop-impl}

In practice, we do not want the sizes of the sets $(A_k, B_k)$ in the iteration to grow too large as this slows computation, and 
large bimodules are difficult to interpret. Therefore, the search procedure is terminated when the geometric size of $(A_k, B_k)$ exceeds 5000. 
In some cases, the sequence of iterates $(A_k, B_k)$ for $k \in \{ 1, \ldots, k_{max}\}$ will form a cycle of length greater than 1, and will therefore fail to reach a fixed point. To search for a nearby fixed point instead, when we encounter the cycle $(A_k, B_k) = (A_l, B_l)$ for some $l < k-1$, we set $(A_{l+1}, B_{l+1})$ to $(A_k \cap A_{k-1}, B_k \cap B_{k-1})$ and continue the iteration. 
\subsection{Initialization Heuristics for BSP}
\label{sec:initilzation-heursitics}

In practice, BSP is initialized with each singleton pair $(\{s\}, \emptyset)$ for $s \in S$, and each singleton pair $(\emptyset, \{t\})$ for $t \in T$.  
When either of the sets $S$ or $T$ is large, we use additional strategies to speed up computation.  
When $|S| \gg |T|$, we initialize BSP from all the features in $T$, but only from a subset of randomly chosen features in $S$.

BSP sometimes discovers identical or almost identical bimodules when starting from different initializations, often from features within the said bimodule. 
This problem is particularly prominent for large bimodules which may be rediscovered by thousands of initializations. Hence, to avoid some of this redundant computation, we provide an option to skip initializing BSP from features in the bimodules that have already been discovered. This option was not however used for the examples in this paper.

\subsection{Covariate Correction}
\label{sec:covcorr}

In some cases the data matrices $[\mathbb{X}, \mathbb{Y}] \in \R^{n \times (p+q)}$ are accompanied by one or more covariates like sex, platform details and PEER factors that must be accounted for by removing their effects before discovering bimodules. Suppose we are given $m$ such linearly independent covariates $v_1, \ldots, v_m \in \R^{n}$. Here we describe how to modify BSP to remove their effects. 
First, we residualize each column of the original data $[\mathbb{X}, \mathbb{Y}]$ by setting up a linear model with explanatory variables $v_1, \ldots, v_m$. Denote the resulting matrix by $[\mathbb{X}', \mathbb{Y}'] \in \R^{n \times (p+q)}$ that has columns which are projections of those of $[\mathbb{X}, \mathbb{Y}]$ onto the subspace orthogonal to $v_1, \ldots v_m$. We would like to now run BSP on $[\mathbb{X}', \mathbb{Y}']$, however since the columns of $D' = [\mathbb{X}', \mathbb{Y}']$ lie on an $n' = n - m'$ dimensional subspace, the permutation p-values (Section \ref{sec:perm-null-dist-and-pvals}) based on data $D'$ will tend to be significant even if $\bbX$ and $\bbY$ were generated independently. However, following \cite{fredPerm}, the p-value approximation can be corrected by replacing the sample size $n$ with the effective sample size of $n'$ in the moment calculations.

\subsection{Uniformity of Our P-Value Estimates}
\label{sec:pval-tails}
\begin{figure}
	\centering    
	\includegraphics[height=3in]{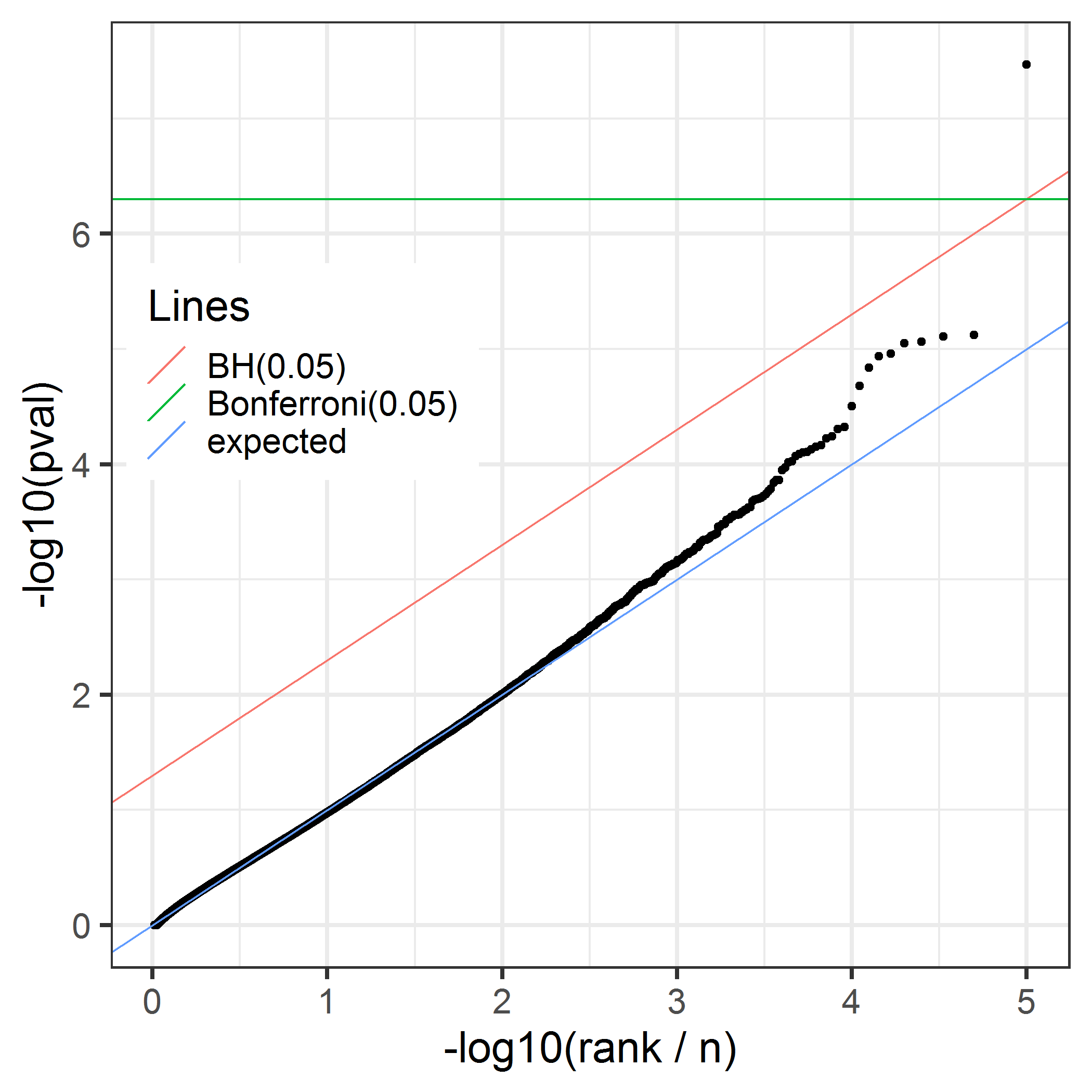}
	\caption{Assessing the accuracy of our p-value estimate $\hat{p}(A,t)$:
		We used the eQTL data from Section \ref{sec:eqtl} and chose a bimodule with 24 SNPs (used as $A$) and selected $t$ to be a gene from the same bimodule. We then performed $10^5$ random permutation of the sample labels for the gene $t$ and repeatedly estimated $\hat{p}(A, t)$ for each permutation after removing the effects of covariates (\smref{sec:covcorr}).}
	\label{fig:p-value-approx}
\end{figure}

For a quick check of the uniformity of our p-value approximation under the permutation null, we chose a bimodule $(A, B)$ found in Section \ref{sec:eqtl} and a $t \in B$. Then we randomly permuted the labels of gene $t$ ($10^5$ times), computing our estimate $\hat{p}(A,t)$ of the permutation p-value $p(A,t)$ (Section \ref{sec:perm-null-dist-and-pvals}) in each case. Hence we are assessing the uniformity of $\hat{p}(A,t)$ under the permutation null distribution. The result in Figure \ref{fig:p-value-approx} shows that the computed p-values are almost uniform but extremely small p-values show anti-conservative behavior. A potential reason for this anti-conservative behavior is that the tails of test statistic under the permutation distribution may be heavier compared to the tails of the location-shifted Gamma distribution that we use to approximate it, since the permutation distribution is a discrete distribution which explicitly depends on the exact entries of the data matrices.

\section{Simulation Details}
\label{sec:sim-appendix}
\subsection{Theory to Justify Simulation Parameter Choice}
\label{sec:sim-details-appendix}

\begin{lemma}
	\label{lem:regressor}
	Fix $\rho, \eta \in [0,1]$, $a, b \in \nat$, and $d \in \{1, 2 \ldots a\}$ so that $\delta \doteq 1 + \rho(d-1) \geq \eta^2 d$. Suppose $\bX$ is an $a$-dimensional random column vector with covariance matrix $\Cov(\bX) = \rho U_a + (1-\rho) I_a$, where $U_a \in \R^{a \times a}$ is the matrix of all ones and $I_a \in \R^{a \times a}$ is the identity matrix. Next suppose $D$ is a $\{ 0,1\}$ valued $a \times b$ dimensional matrix that has exactly $d$ ones in each column. Finally, let $\sigma = \sqrt{\delta(\delta - \eta^2d)}/\eta$ and suppose the $b$-dimensional random column vector $\bY$ is given by
	\begin{equation*}
		\bY = D^t \bX + \boldsymbol{\epsilon}
	\end{equation*}
	where $\boldsymbol{\epsilon}$ is another $b$-dimensional random vector independent of $\bX$ with $\Cov(\boldsymbol{\epsilon}) = \sigma^2 I_b$. Then 
	\begin{equation}
		\label{eq:correlation-on-edges}
		\Cor(\bX, \bY) \odot D = \eta D
	\end{equation}
	where $\Cor(\bX, \bY) \in \R^{a \times b}$ is the cross-correlation matrix between random vectors $\bX$ and $\bY$, and $\odot$ represents the element-wise product of matrices (i.e., the Hadamard product).
\end{lemma}
\begin{proof}
	Since we are concerned with covariances, we can assume by mean centering that $\Ex \bX = 0 \in \R^a$ and $\Ex \bY = \Ex \epsilon = 0 \in \R^b$. Note that $D^t e_a = d e_b$ and $U_a = e_ae_a^t$, where $e_r \doteq (1, \ldots, 1)^t \in \R^r$ for $r \in \{a,b\}$ is the $r$-dimensional vector with all entries equal to one. Using independence of $\bX$ and $\boldsymbol{\epsilon}$:
	\begin{equation*}
		\begin{aligned}
			\Cov(\bY) = \Ex (\bY\bY^t) &= D^t \Ex(\bX \bX^t) D + \Ex (\boldsymbol{\epsilon \epsilon^t})  \\
			&= D^t \Cov(\bX) D + \Cov(\boldsymbol{\epsilon}) = D^t (\rho e_ae_a^t + (1-\rho)I_a) D + \sigma^2 I_b\\
			&= \rho (D^t e_a) (D^t e_a)^t + (1-\rho) D^t D + \sigma^2 I_b \\
			&= \rho d^2 e_be^t_b + (1-\rho)   D^t D + \sigma^2 I_b.
		\end{aligned}
	\end{equation*}
	Since all the diagonal entries of $D^t D$ have the value $d$,
	\begin{equation}
		\label{eq:variance}
		\diag{\Cov(\bY)} = (\rho d^2 + (1-\rho) d + \sigma^2) I_b = (d \delta  + \sigma^2) I_b = \brR*{\frac{\delta}{\eta}}^2 I_b
	\end{equation} 
	where for any square matrix $A$, $\diag{A}$ denotes the diagonal matrix obtained from $A$ by setting all the off-diagonal entries of $A$ to $0$.
	
	We can similarly calculate the cross-covariance between $\bX$ and $\bY$
	\begin{equation}
		\begin{aligned}
			\Cov(\bX, \bY) = \Ex (\bX \bY^t) &= \Ex (\bX\bX^t)D = (\rho e_ae_a^t + (1-\rho) I_a) D \\
			&=  \rho d e_ae_b^t + (1-\rho) D.  
		\end{aligned}
		\label{eq:corr-cov}
	\end{equation}
	Thus, using  \eqref{eq:variance}, \eqref{eq:corr-cov}, and $\diag{\Cov(\bX)} = I_a$, the cross-correlation between $\bX$ and $\bY$ is equal to:
	\begin{equation*}
		\begin{aligned}
			\Cor(\bX, \bY) &= \diag{\Cov(\bX)}^{-\frac{1}{2}} \Cov(\bX, \bY) \diag{\Cov(\bY)}^{-\frac{1}{2}} \\
			&= \frac{\eta}{\delta}\brR*{\rho d e_ae_b^t + (1-\rho) D} = \frac{\eta}{\delta}\brR*{\rho d \bar{D} + (1-\rho + \rho d) D}\\
			&= \eta D + \eta \rho d \delta^{-1} \bar{D}.  
		\end{aligned}
	\end{equation*}
	where $\bar{D} \doteq e_ae_b^t - D$ is the complement of $D$, i.e. $D_{ij} = 1-\bar{D}_{ij}$ for $i, j$. In particular this shows \eqref{eq:correlation-on-edges}.
\end{proof}

\subsection{Running BSP and Related Methods}
\label{sec:sim-results}

We applied BSP to the simulated data using the false discovery parameter $\alpha = 0.02$, which was selected to 
keep the edge-error estimates under $0.05$ (see Section \ref{sec:howtochoosealpha}). 
This tuning procedure is purely based on the observed data, and does not use any knowledge of the ground truth.
The search was initialized from 
singletons consisting of all the features in $T \cup S$. 
In what follows, feature set pairs identified by BSP (or some other method, when clear from context) will
be referred to as {\em detected} bimodules. BSP detected 708 unique bimodules while the effective number (see Section \ref{sec:post-processing}) of detected bimodules was 644.87.

To obtain bimodules via CONDOR \citep{platig2016bipartite}, we first applied \meqtl\ \citep{matrixEQTL} 
to the simulated dataset  with $S$ considered as the set of SNPs and $T$ considered as the set of genes, to extract feature pairs $(s,t) \in S \times T$ with q-value less than $0.05$.
Next, we formed a bipartite graph on the vertex set $S \cup T$ with edges given by 436,616 significant feature pairs 
found by \meqtl. The largest connected component of this graph, made up of 48,455 features from $S$ and 11,045 features from $T$, was passed through a bipartite community detection software \citep{condor-software} which partitioned the nodes of the subgraph into 178 bimodules.

We applied the sCCA method of \cite{witten2009penalized} to the simulated data to find $100$ bimodules.
More precisely, for various penalty parameters $\lambda \in [0,1]$, we ran sCCA \citep{pma} to find $100$ canonical covariate pairs with the $\ell_1$ norm constraint of $\lambda \sqrt{p}$ and $\lambda \sqrt{q}$ on the coefficients of the linear combinations corresponding to $S$ and $T$ respectively. Initially, we considered  $\lambda = 0.1$, chosen by the permutation based procedure provided with the software. However, the resulting bimodules were very large and had high edge-error (further details are provided in Section \ref{sec:scca-results}). Based on a rough grid search, we then ran the procedure with each value $\lambda \in \{.01,.02,.03,.04 \}$ to obtain smaller bimodules.

\subsubsection{Comparing Performance of the Methods}
\label{sec:sim-performance}
In the simulation study described above, we measure the recovery of a target bimodule $(A_t, B_t)$ by a detected bimodule $(A_d,B_d)$ using the two metrics:
\[
\mbox{recall} = \frac{|A_t \cap A_d||B_t \cap B_d|}{|A_t||B_t|} 
\ \ \mbox{ and } \ \ 
\mbox{Jaccard} = \frac{|A_t \cap A_d||B_t \cap B_d|}{|(A_t \times B_t) \cup (A_d \times B_d)|}. 
\]
Recall captures how well the target bimodule is \emph{contained} inside the detected bimodule, while Jaccard measures how well the two bimodules \emph{match}. When assessing the recovery of a target bimodule under a collection of detected bimodules (like the output of BSP), we choose the detected bimodule with the best recall or Jaccard, depending on the metric under consideration.

As shown in Figure \ref{fig:recall}, the  BSP Jaccard for target bimodules was influenced primarily by 
the cross-correlation strength $\sqrt{\frac{r^2(A,B)}{|A||B|}}$ of the target bimodule, though the intra-correlation parameter $\rho$ 
used in the simulation \eqref{eq:reg} was also seen to have an effect (Figure \ref{fig:recall}, left). 
Most bimodules with cross-correlation strength above 0.4 were completely recovered, 
while those with strength below 0.2 were not recovered. 
For strengths between 0.2 to 0.4, there was a variation in Jaccard, with smaller Jaccard for bimodules 
having larger values of $\rho$ (Figure \ref{fig:recall}, left). 
The effect of $\rho$ on Jaccard was expected since BSP accounts for the intra-correlation among features of the same type. 

The intra-correlation parameter $\rho$ did not have significant effect on CONDOR Jaccard, since the method does not account for intra-correlations. Hence, here we only consider the effects of the cross-correlation strength of target bimodules on CONDOR Jaccard 
(Figure \ref{fig:recall}, right). Regardless of the cross-correlation strength, CONDOR Jaccard remained low. This was because CONDOR bimodules often overlapped with multiple target bimodules; indeed, 155 of the 178 CONDOR bimodules overlapped with two or more (up to 14) target bimodules, compared with only 58 of the 708 BSP bimodules.
However, the results for CONDOR recall (Figure \ref{fig:recall}, right) show that most 
target bimodules with significant cross-correlation strengths were contained inside some CONDOR bimodule.

To assess the false discoveries in detected bimodules, we measured the \emph{edge-error} 
of detected bimodules.  The edge-error is the fraction of the essential-edges (Definition \ref{def:ess-edges}) 
of a detected bimodule that are not part of the simulation model, that is, edges not contained in
any target bimodule and not in the set of bridge edges. 
The average edge-error for BSP bimodules was 0.041, and 90\% of the detected bimodules had edge-error under 0.11. 
In contrast, the average edge-error for CONDOR bimodules was 0.09, and 90\% of the detected bimodules had 
edge-error under 0.20. The larger edge-error among CONDOR bimodules may have arisen because the method 
does not account for intra-correlations.

Concerning sCCA, the sizes of the detected bimodules were at least an order of 
magnitude larger than sizes of the target bimodules when $\lambda$ exceeded $0.04$ (see Figure \ref{fig:scca-sizes} in Section  \ref{sec:scca-results}). 
Thus we only considered $\lambda \leq 0.04$. For $\lambda = 0.02$, $0.03$, and $0.04$, the detected bimodules had large edge-error 
(average error 0.47, 0.76 and 0.89, respectively), while for $\lambda =  0.01$ the target bimodules had poor 
recall (99\% of the target bimodules had recall below 0.1). Further details of these results are given in  Section \ref{sec:scca-results}. 

A potential shortcoming of our application of sCCA was that we chose the same penalty parameter $\lambda$ for each of the 100 bimodules. 
We expect that the results of sCCA would improve if one chose a different penalty parameter for each bimodule. However, \cite{witten2009penalized} does not provide explicit guidelines to choose different penalty parameters for each component (bimodule), and directly doing a permutation-based grid search each time would be exceedingly slow.

\subsubsection{Results From sCCA} 
\label{sec:scca-results}
\begin{figure}
	\centering
	\includegraphics[width=0.7\textwidth]{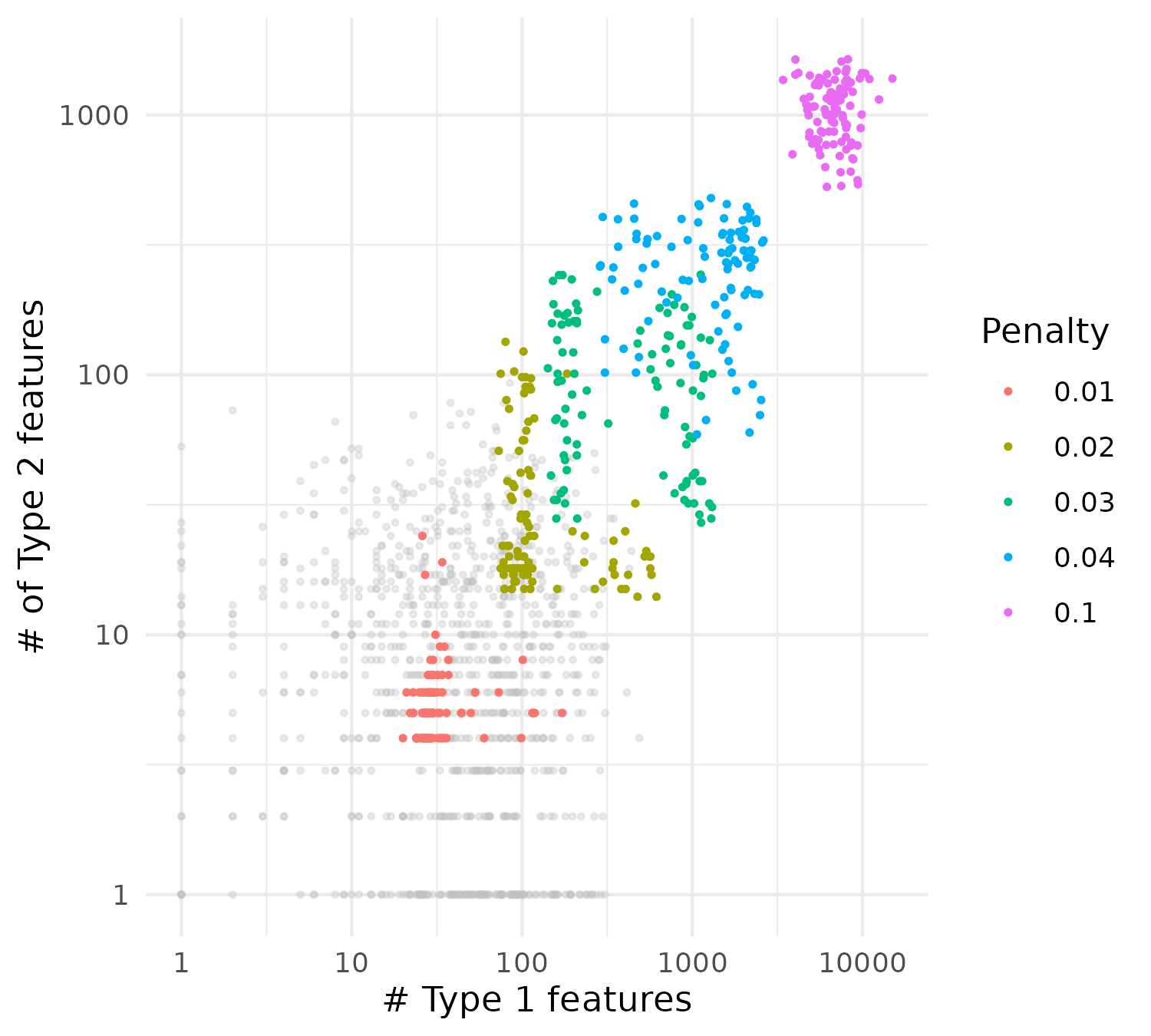}
	\caption{The sizes of sCCA bimodules for various values of the penalty parameter $\lambda$, along with sizes of the target bimodules in gray.}
	\label{fig:scca-sizes}
\end{figure}

As described earlier, we ran sCCA on the simulated data to search for $100$ canonical covariates for a range of values of the penalty parameter $\lambda$. 
The sizes of the bimodules for various values of $\lambda$ can be seen in Figure \ref{fig:scca-sizes}. 
For $\lambda \in \{ 0.01, 0.02, 0.03, 0.04, 0.1\}$, the first two columns of the following table show the number of target bimodules (TB) that overlapped with each detected bimodule (DB) and the edge-error of each DB, both averaged over all DBs.
The last column shows the top 1 (or bottom 99) percentile \textit{recall} among the target bimodules.

\begin{center}
	\begin{tabular}{|c|c|c|c|} \hline
		$\lambda$ & \# TBs that overlap with each DB  & edge-error & recall of TB (99\%-tile)\\
		\hline
		.01 &  .93 & 0.24 & 0.1\\
		.02 &  1.04 & 0.47 & 0.98\\
		.03 & 5.69 & 0.76 & 1\\
		.04 & 24.88 & 0.89 & 1\\
		.1 & 156.28 & 0.94 & 1 \\
		\hline
	\end{tabular}
\end{center}

The parameter value $\lambda = 0.01$ has small edge-error, but poor recall. The recall improves on increasing $\lambda$, but the edge-error degrades.

\subsection{Performance of BSP and CONDOR on Increasing Sample Size}
\label{sec:large-sample-sim}

\begin{figure}
	\centering
	\includegraphics[width=0.5\textwidth]{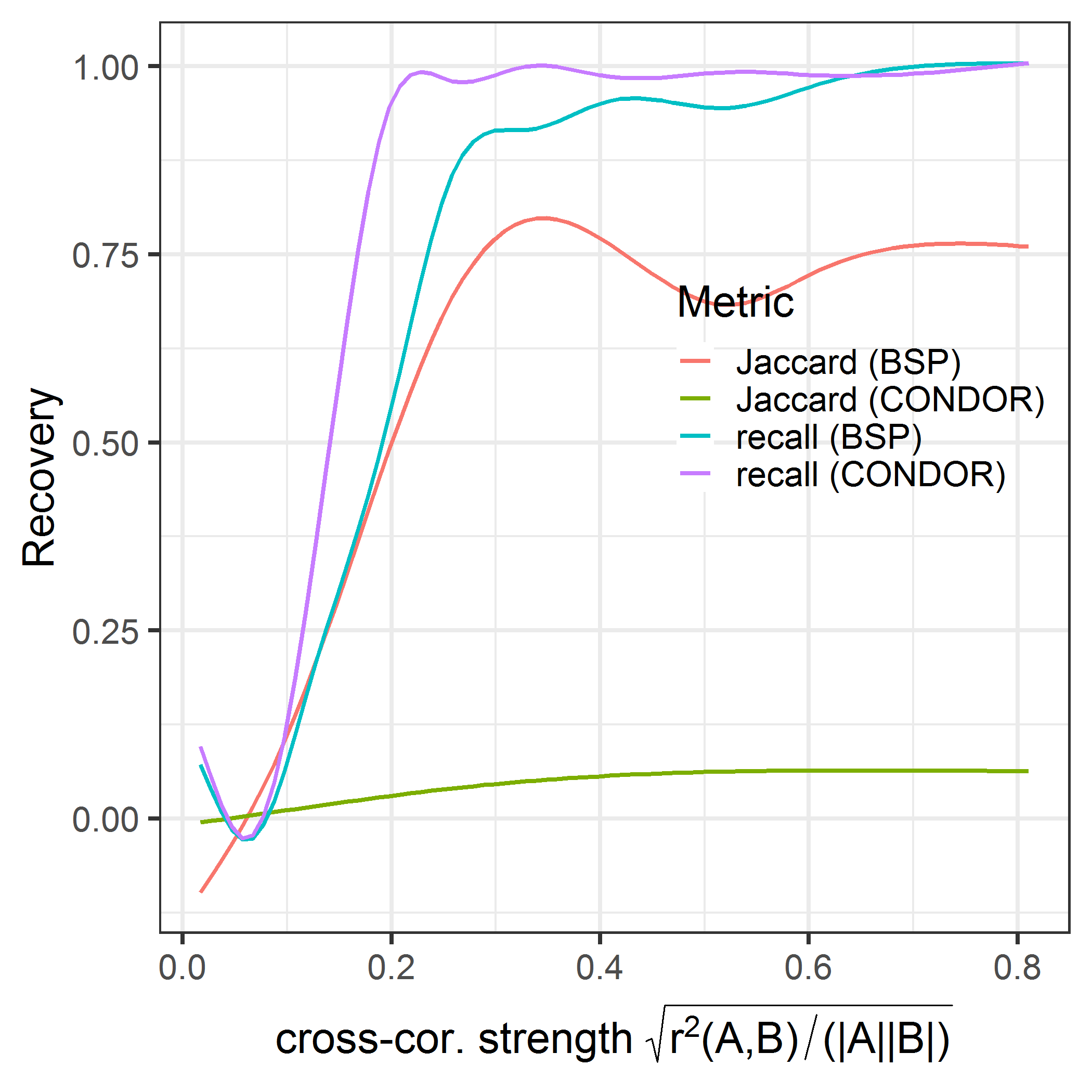}
	\caption{Average recall and Jaccard for target bimodules in the simulation with $600$ samples.}
	\label{fig:large-ss}
\end{figure}

We also increased that sample size of our simulation study to $n=600$, and re-ran BSP and CONDOR with the same parameters as earlier. The  average edge-error for BSP and CONDOR was 0.05 and 0.10 respectively. As seen in Figure \ref{fig:large-ss}, based on  recall, BSP and CONDOR both recover most bimodules with cross-correlation strength above 0.3, however Jaccard for BSP and CONDOR has degraded. This can be explained by noting that 25\% of BSP bimodules now overlapped with two or more target bimodules compared to 8\% when $n=200$.

\section{GTEx Results}
\label{sec:gtex-appendix}
\subsection{Data Acquisition and Preprocessing}
\label{sec:acq-and-preprocessing}

We obtained genotype and thyroid expression data for 574 individuals from the dbGap website (accession number: phs000424.v8.p1). 
We directly used the filtered and normalized gene expression data and covariates provided for eQTL analysis but filtered the SNPs in the genotype data using the LD pruning software \textit{SNPRelate} \citep{zheng2015tutorial}. The software retained 556K autosomal SNPs with minor allele frequency above $0.1$ such that all pairs of SNPs within each 500KB window of the genome had squared correlation under $(0.7)^2$. The latter threshold was chosen to balance the number of retained SNPs and information loss. As SNPs exhibit local correlation due to linkage disequilibrium (LD), the selection process should not reduce the statistical power of BSP.

 There were 68 covariates provided for the Thyroid tissue consisting of the top 5 genotype principal 
components; 60 PEER covariates, and 3 additional covariates for sequencing platform, 
sequencing protocol, and sex. We accounted for these covariates by the modification to BSP mentioned in \smref{sec:covcorr}.

\subsection{Running BSP}
\label{sec:details-of-running-bsp-on-gtex}

We applied BSP to the thyroid eQTL data with false discovery parameter $\alpha = 0.03$ 
selected to keep the edge-error estimates under $0.05$. The search was initialized from singleton sets 
of all genes and half of the available SNPs, chosen at random.  Thus the search procedure in 
Section \ref{sec:bsp} was run $p/2 + q \sim$ 304K times.  
BSP took 4.7 hours to run on a computer with a 20-core 2.4 GHz processor (further processor details are provided 
in \smref{sec:machine-details}). 
The search identified 3744 unique bimodules with p-values below the significance threshold of 
$\frac{\alpha}{pq} = 3.45 \times 10^{-12}$ (see Section \ref{sec:bsp}). The majority (277K) 
of the searches terminated in the empty set after the first step; 
of the remaining 27K searches, the great majority identified a non-empty fixed point within 20 steps. 
Only 20 searches cycled and did not terminate in a fixed point. 
Among the searches taking more than one iteration, 
94\% terminated by the fifth step.  Among searches that found a non-empty fixed point, 92.3\% of the fixed points
contained the seed singleton set of the search.

The effective number (see Section \ref{sec:post-processing}) of bimodules was 3304, slightly smaller than the number of unique bimodules. 
We applied the filtering procedure described in Section \ref{sec:post-processing} to select from the unique bimodules 
a subfamily of 3304 bimodules that were substantially disjoint. 
The selected bimodules had SNP sets ranging in size from 1 to 1000, and gene sets ranging in size 
from 1 to 100 (Figure \ref{fig:size}); the median size of the gene and SNP sets was 1 and 7, respectively.

If required, BSP can be run in a faster (less exhaustive) or slower (more exhaustive) fashion by selecting a smaller 
or larger fraction of SNPs from which to initialize the search procedure. 
The effective number of discovered bimodules was only slightly smaller (3258) when initializing with 10\% of the SNPs.

\subsection{Running Other Methods}
\label{sec:gtex-running-other-methods}

Standard eQTL analysis was performed by applying Matrix-eQTL \citep{matrixEQTL} twice to the data, 
first to perform a \ciseqtl{} analysis within a distance of 1MB and next to perform a \transeqtl{} analysis. 
In each case, SNP-gene pairs with BH $q$-value less than $0.05$ were identified as significant. 
Matrix-eQTL identified 186K \ciseqtl{}s and 73K \transeqtl{}s.

To obtain CONDOR bimodules \citep{platig2016bipartite}, we applied Matrix-eQTL to identify both \textit{cis-} and \textit{trans-}eQTLs with BH q-value under the threshold $.2$, chosen as in \cite{fagny2017exploring}. 
The resulting gene-SNP bipartite graph formed by these eQTLs was passed through CONDOR's bipartite
community detection pipeline \citep{platig2016bipartite}, which partitioned the nodes of the largest connected component of this graph 
into 6 bimodules.

We also applied the sCCA method of \cite{witten2009penalized} using the permutation based parameter selection procedure \citep{pma} on the covariate-corrected genotype and expression matrices to identify $50$ 
bimodules. 
The identified bimodules were large, containing roughly 100K SNPs and 4K-8K genes (Figure \ref{fig:size}), 
making them difficult to analyze and interpret.  The identified bimodules also exhibited moderate overlap: the effective 
number was 25.  As such, we excluded the sCCA bimodules from subsequent comparisons.
Analysis of sCCA on the simulated data (Section \ref{sec:sim-performance}) suggests that the method may be able to recover smaller bimodules with a more tailored choice of its parameters.  

\subsection{Choice of BSP parameter $\alpha$}

\label{sec:gtex-choice-of-alpha}

We chose the false discover parameter $\alpha$ for BSP from the grid $\{0.01, 0.02, 0.03, 0.04, 0.05\}$ by finding the largest $\alpha$ that kept the average edge-error estimates based on $N=5$ half-permutations under 0.05 (Section \ref{sec:howtochoosealpha}). However our error estimates were variable as we obtained $\alpha = 0.05$ in one instance and $\alpha = 0.03$ in another. We conservatively chose $\alpha = 0.03$. 

\subsection{Hardware and Software Stack}
\label{sec:machine-details}

The various methods used is this analysis were run on a dedicated computer that had Intel (R)  Xeon (R) E5-2640 CPU with 20 parallel cores at 2.50 Hz base frequency, and a 512 GB random access memory along with L1, L2 and L3 caches of sizes 1.3, 5 and 50 MB respectively. The computer ran Windows server 2012 R2 operating system and we used the Microsoft R Open 3.5.3 software to perform most of our analysis, since it has multicore implementations of linear algebra routines.

\subsection{Bimodule Connectivity Thresholds and Network Sparsity}
\label{sec:network-plot-details}

\begin{figure}
\centering
 \includegraphics[scale=0.8]{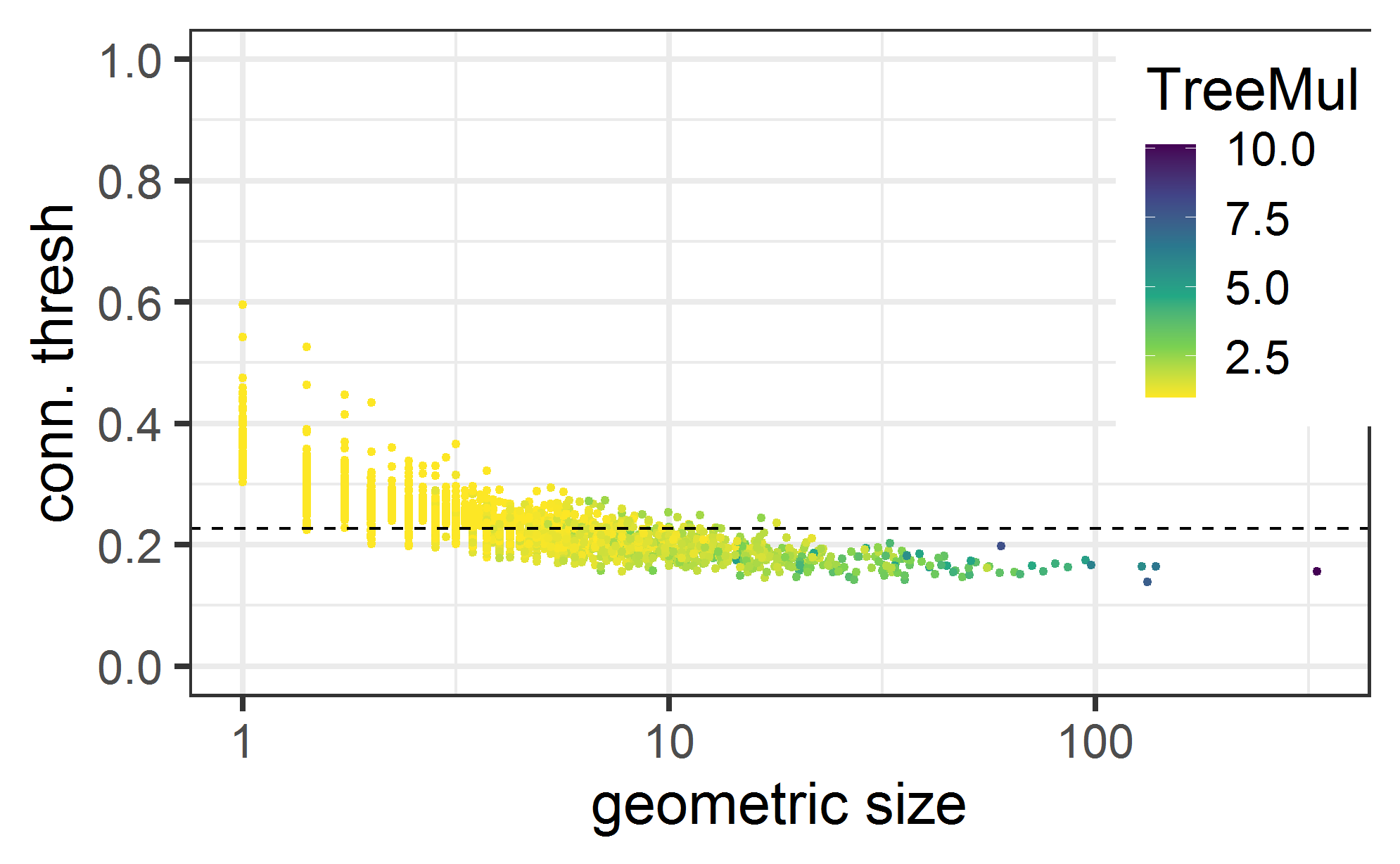}
 \caption{Connectivity-threshold and tree-multiplicity for BSP bimodules compared to their geometric size. The horizontal dotted line represents the threshold obtained from standard trans-analysis.}
 \label{fig:network}
\end{figure}

Figure \ref{fig:network} shows two network statistics for bimodules found by BSP: connectivity threshold and tree-multiplicity. All bimodules have tree multiplicity under $10$. This shows that the association network for large bimodules, particularly having low connectivity-thresholds, is sparse. 

\subsection{Connectivity of Bimodules Under Edges From Combined eQTL Analysis}
\label{sec:bimodule-eqtl-connectivity}

\begin{figure}
\centering
\begin{subfigure}{.5\textwidth}
  \centering
 \includegraphics[width=\linewidth]{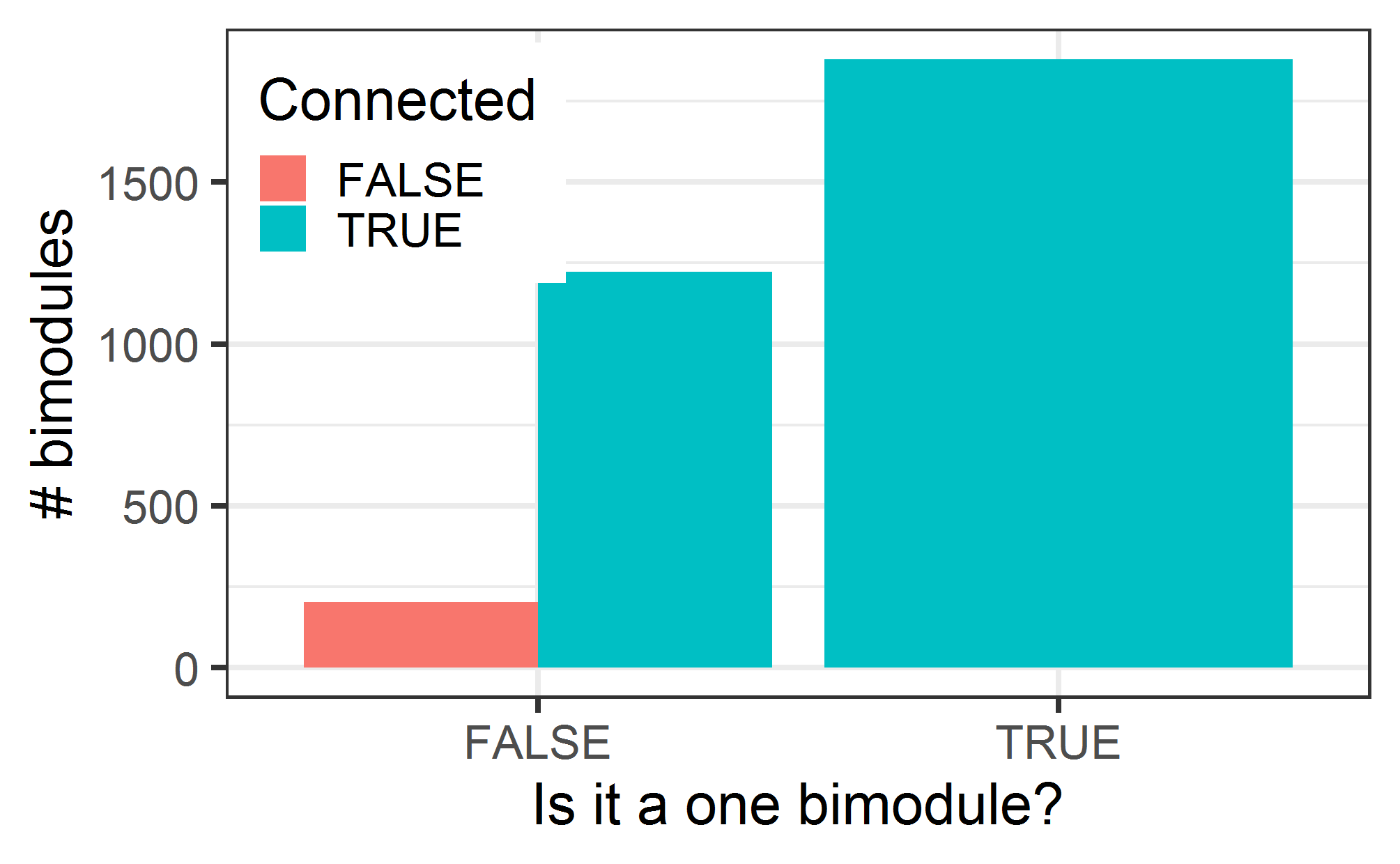}
\end{subfigure}\begin{subfigure}{.5\textwidth}
  \centering
  \includegraphics[width=\linewidth]{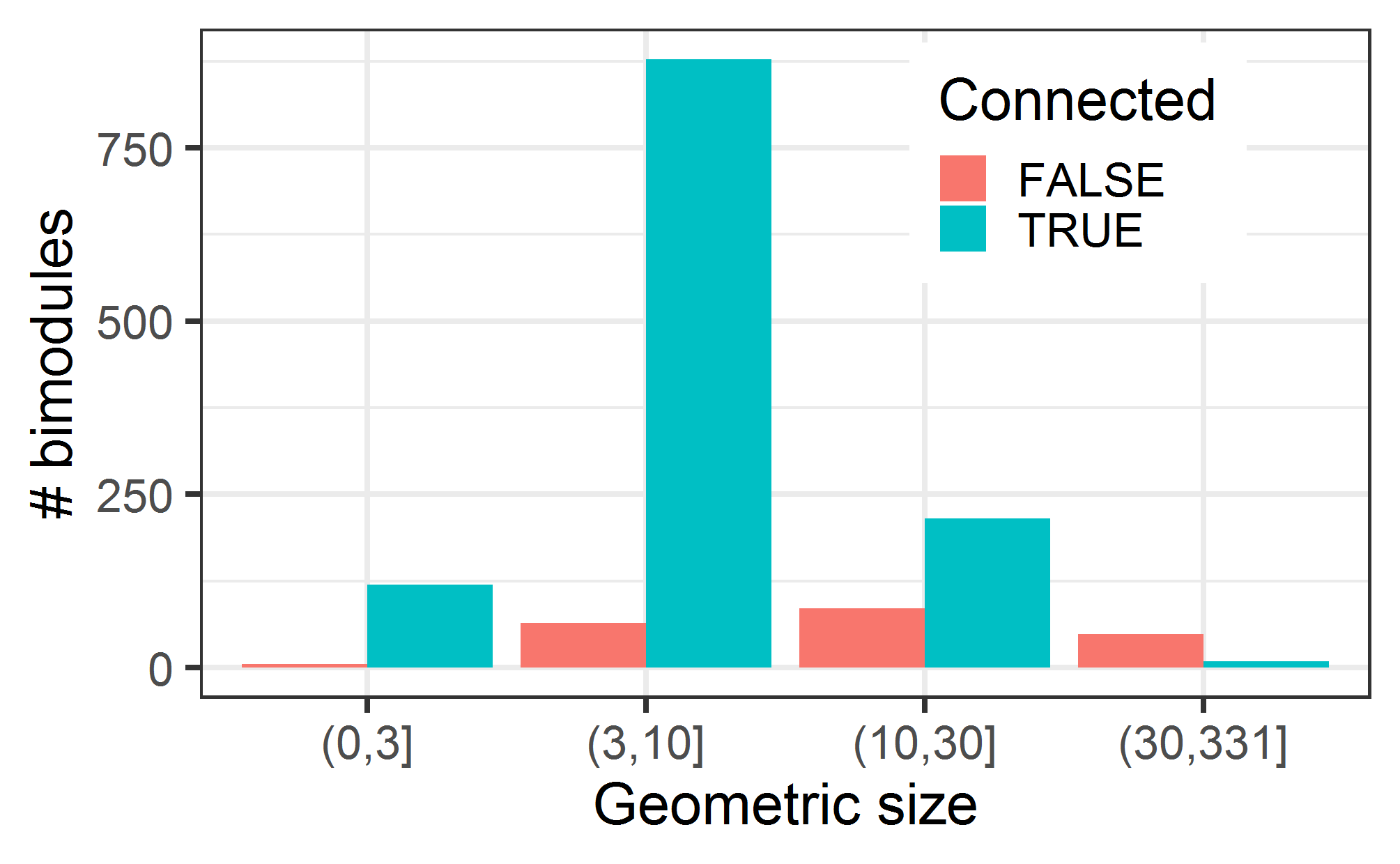}
\end{subfigure}
 \caption{Connectivity of BSP bimodules under combined edges from \ciseqtl{} and \transeqtl{} analysis. Left: the number of bimodules that are connected and are \emph{one bimodules} (i.e.~have one gene or one SNP). Right: Among bimodules having two or more genes and SNPs (i.e. are not one bimodules), the geometric-mean size of the bimodules and their connectivity based on eQTLs from standard analysis.}
\label{fig:eqtl-con}
\end{figure}

Here we examine which bimodules are connected under the combined edges from \ciseqtl{} and \transeqtl{} analysis, based on geometric-mean size of the bimodule. 
Figure   \ref{fig:eqtl-con} (left) shows that all the bimodules that have either one gene or one SNP are connected. Hence, these bimodules could have been recovered using standard eQTL analysis. On the other hand if we restrict to bimodules with two or more genes and SNPs, we see that (Figure \ref{fig:eqtl-con}; right) the fraction of connected bimodules tends to decrease as the geometric-mean size of the bimodules increases.

\newcommand{\bmwidth}{0.33\textwidth}

\subsection{Bimodule Association Networks}
\label{sec:bimod-example-plots}
\begin{figure}
	\begin{tabular}{ccc}
		\includegraphics[width = \bmwidth]{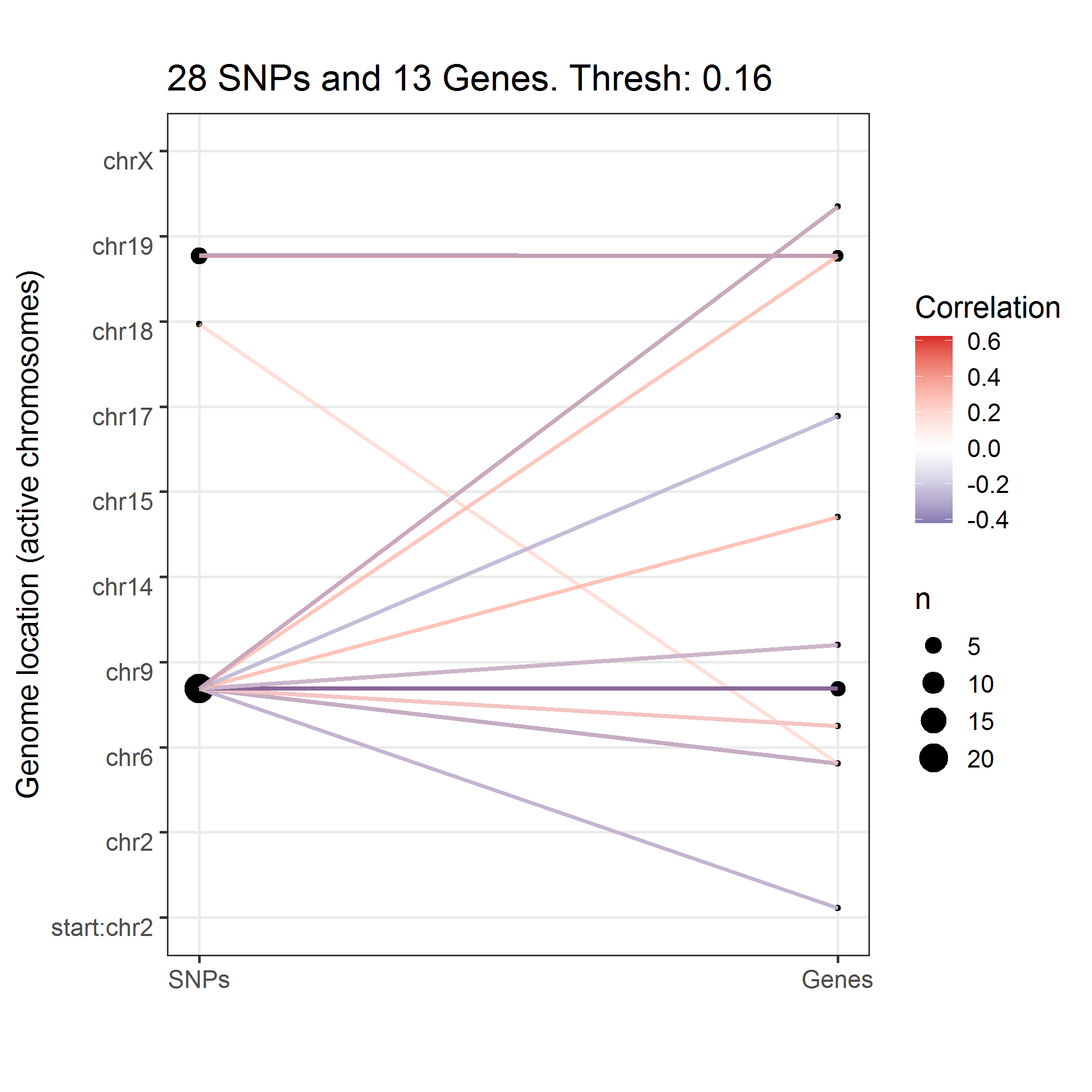} &
		\includegraphics[width = \bmwidth]{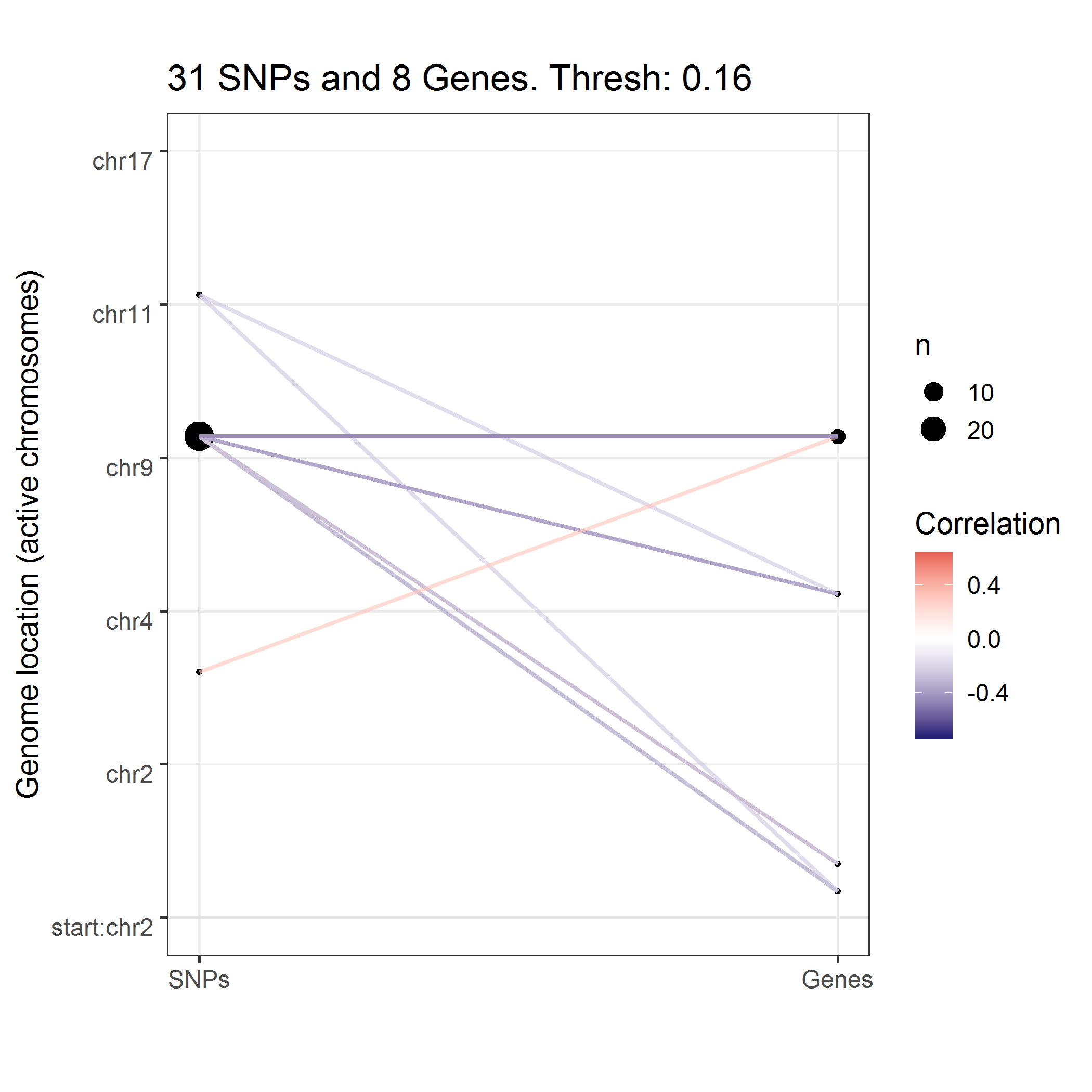} &
	    \includegraphics[width = \bmwidth]{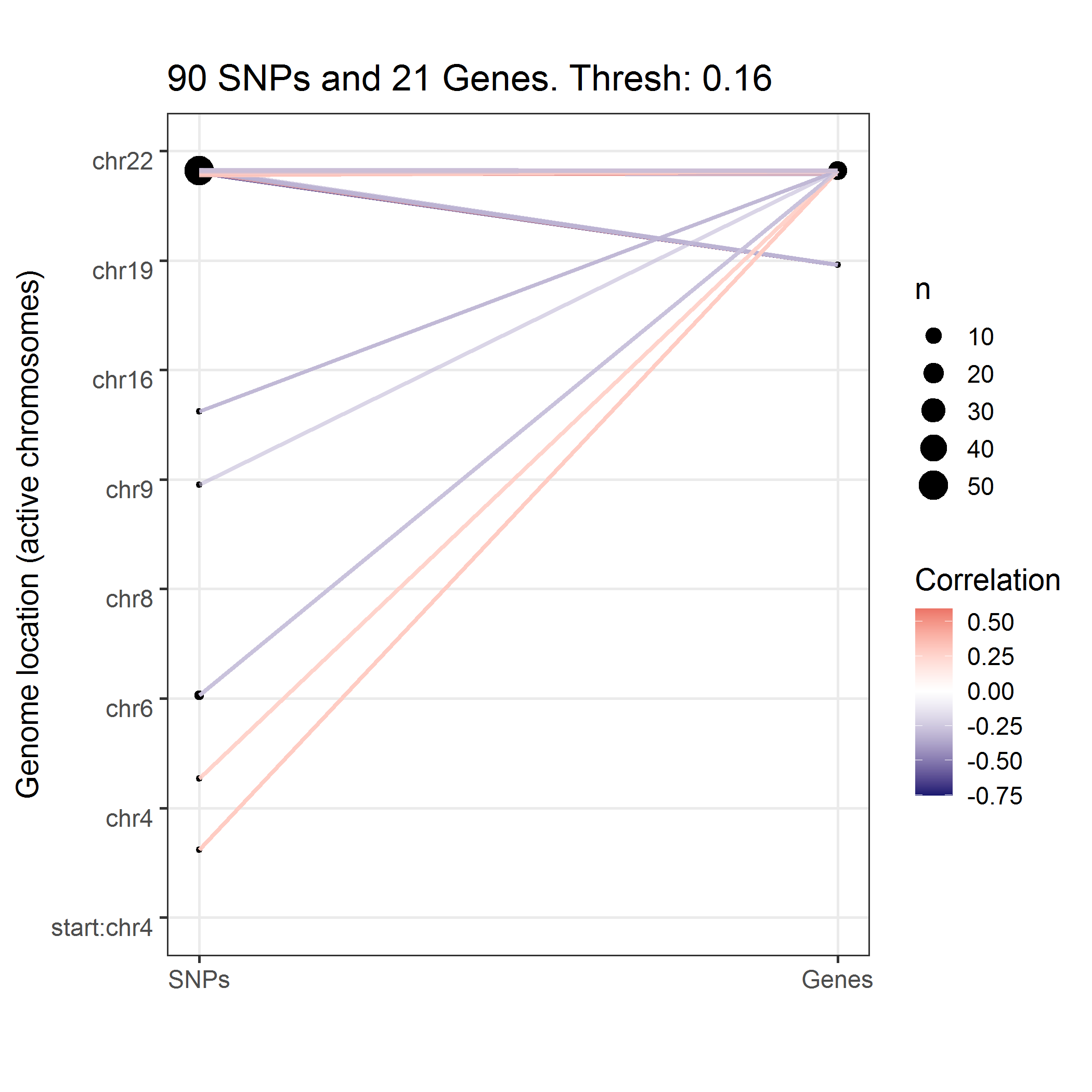} \\
	    \includegraphics[width = \bmwidth]{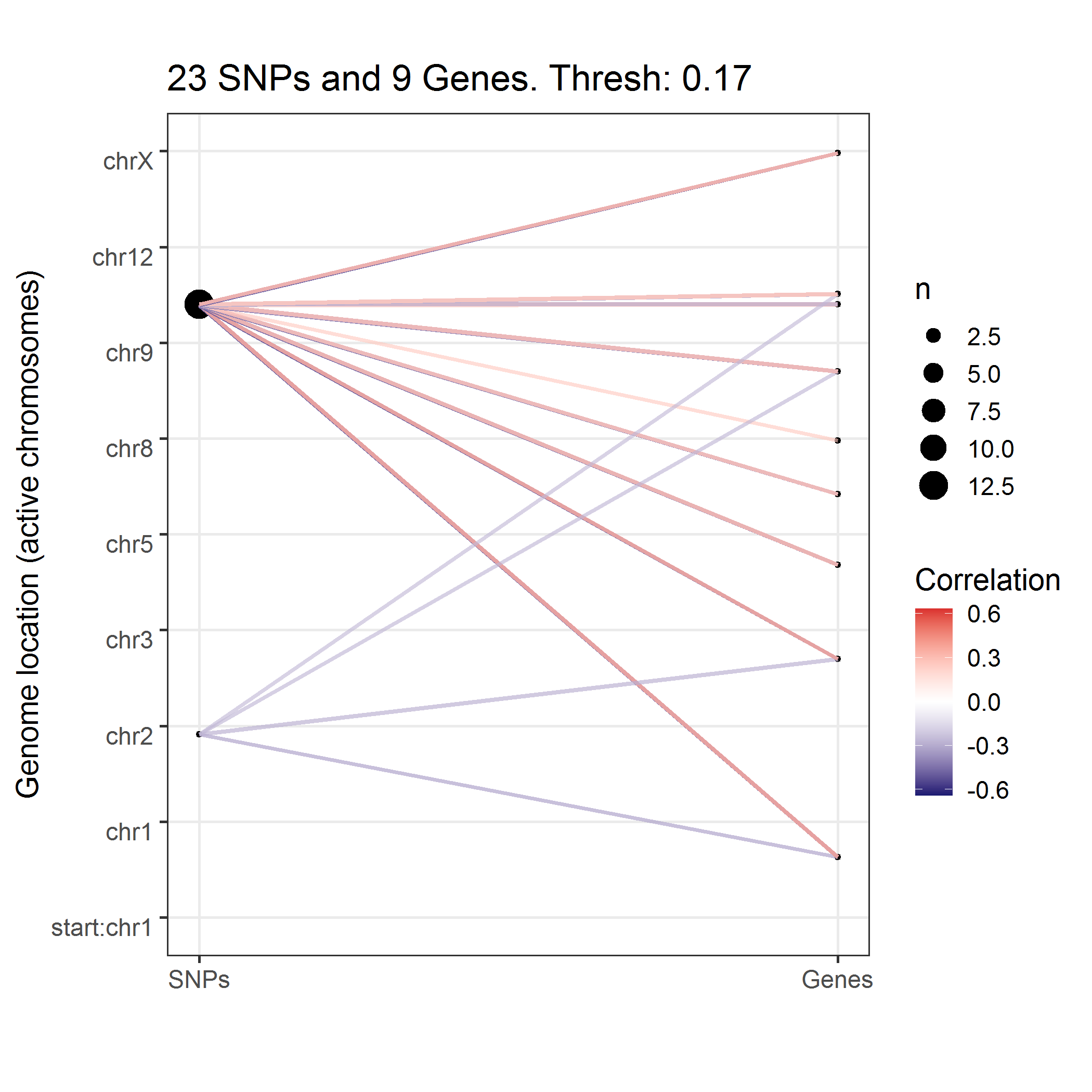} &
	    \includegraphics[width = \bmwidth]{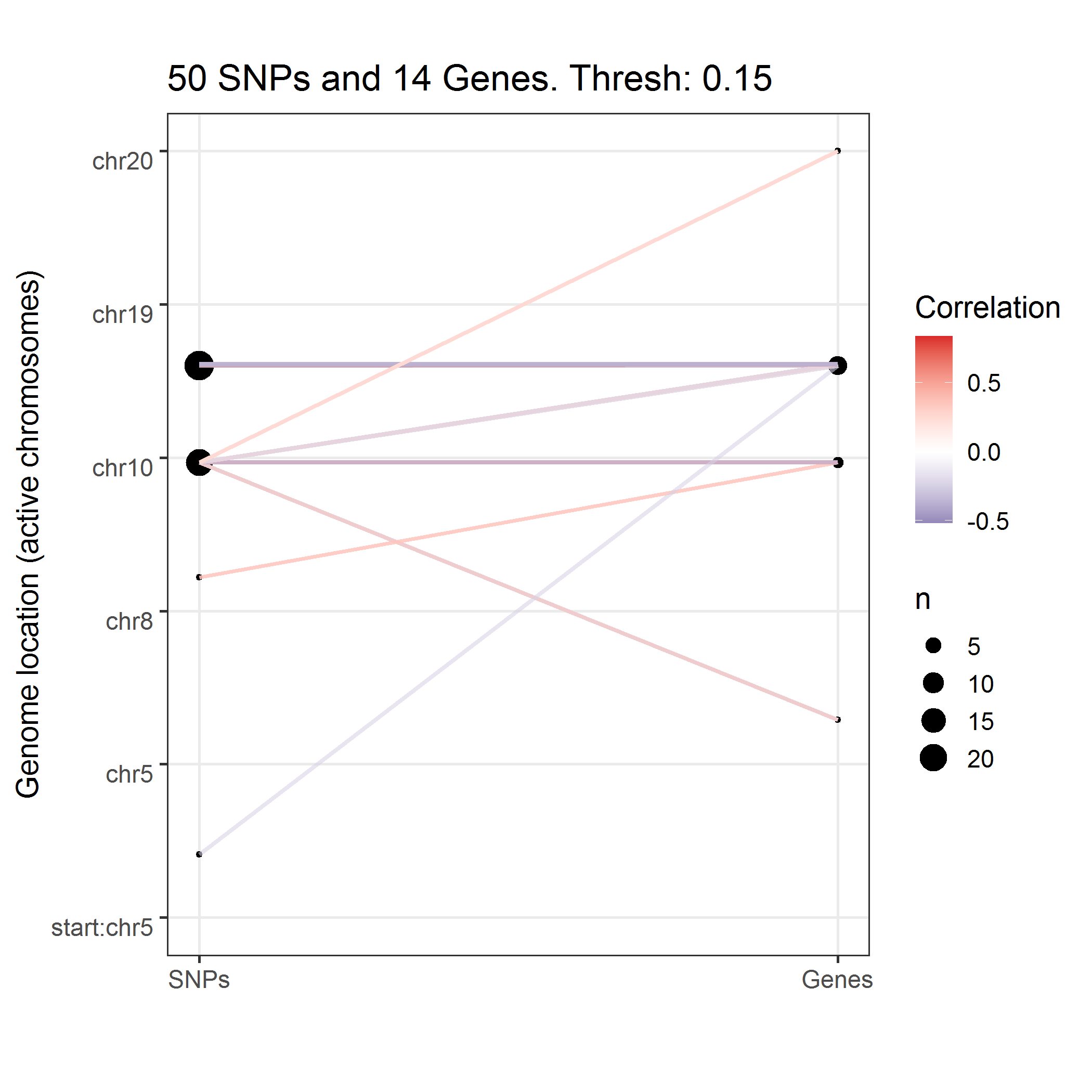} &
	    \includegraphics[width = \bmwidth]{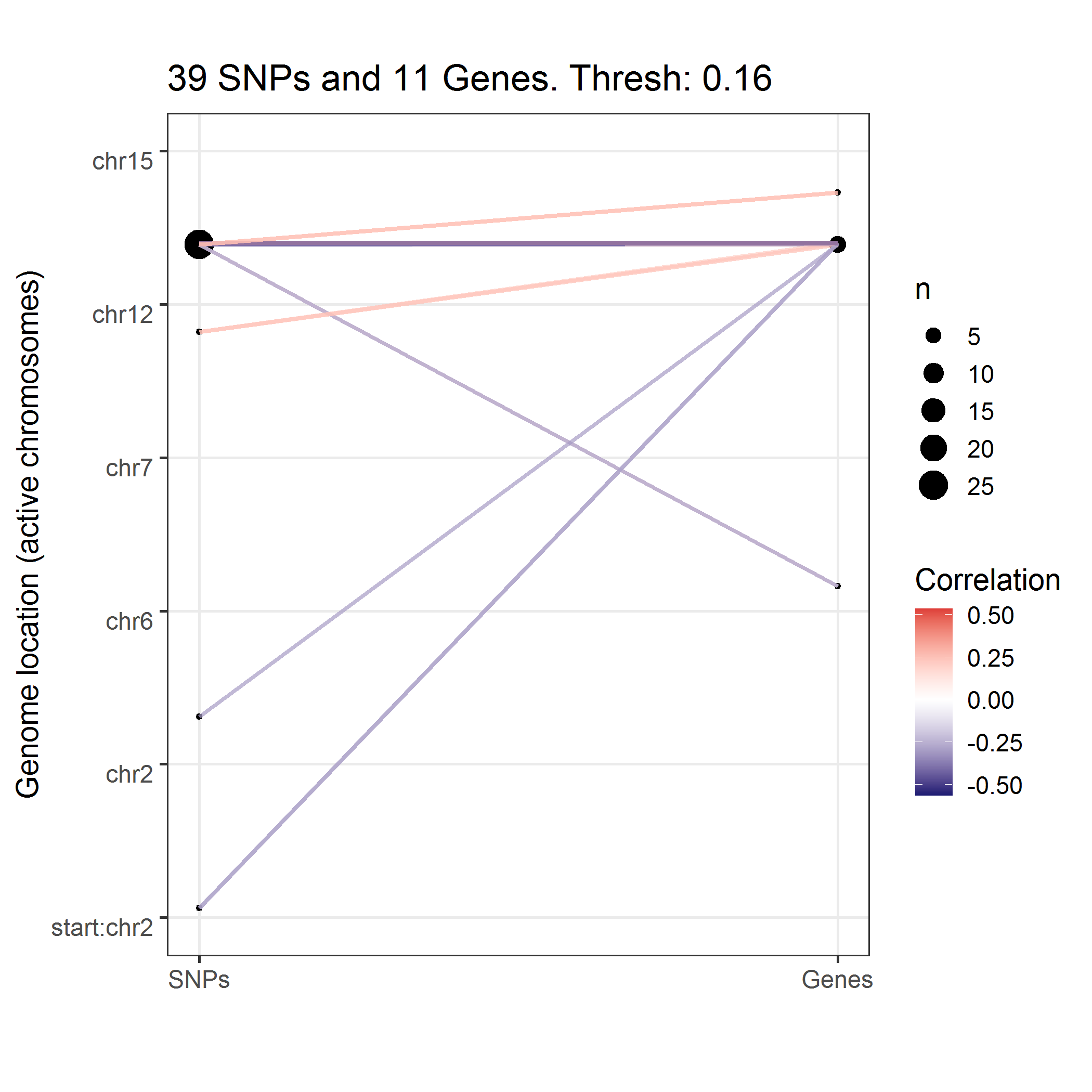} \\
		\includegraphics[width = \bmwidth]{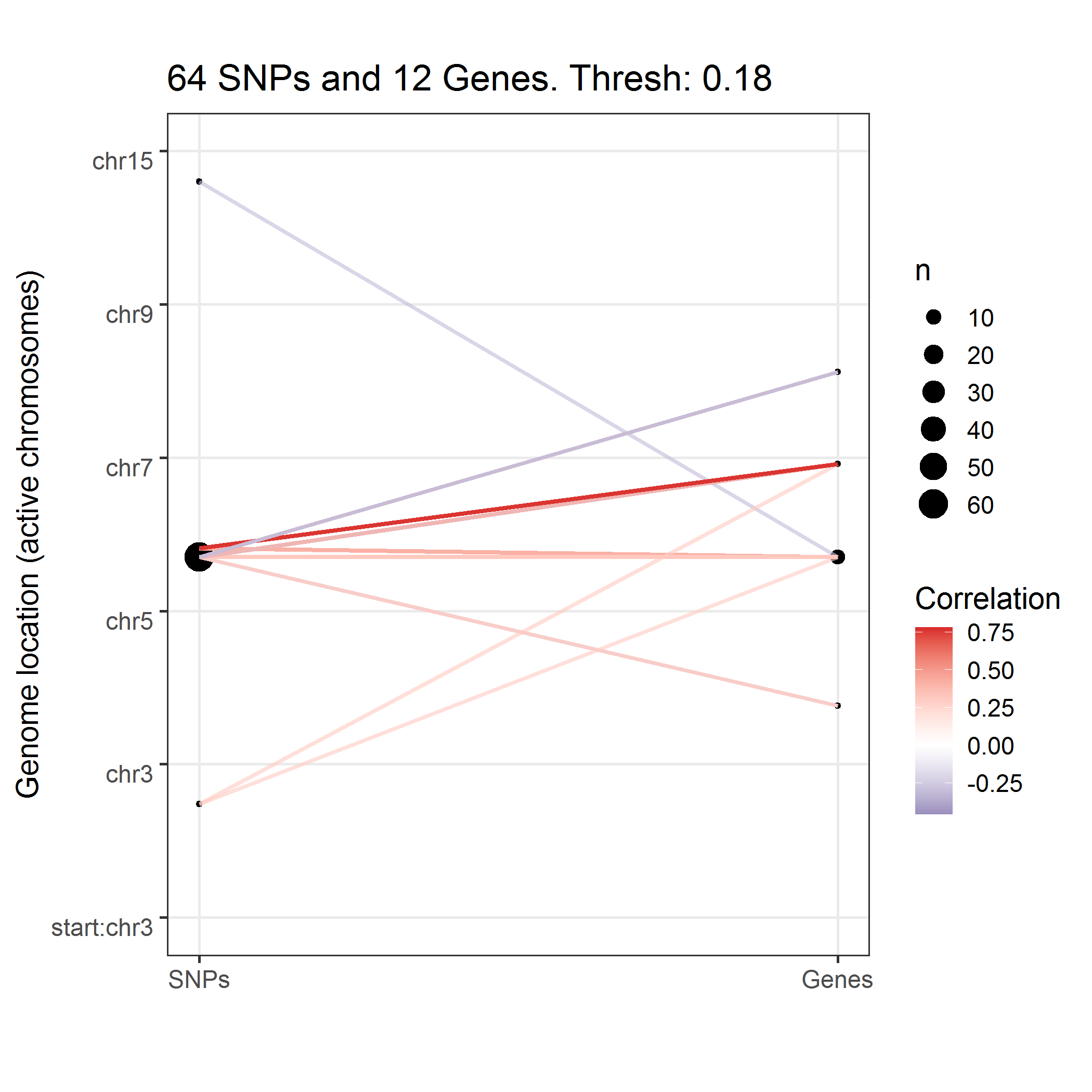}  &
        \includegraphics[width = \bmwidth]{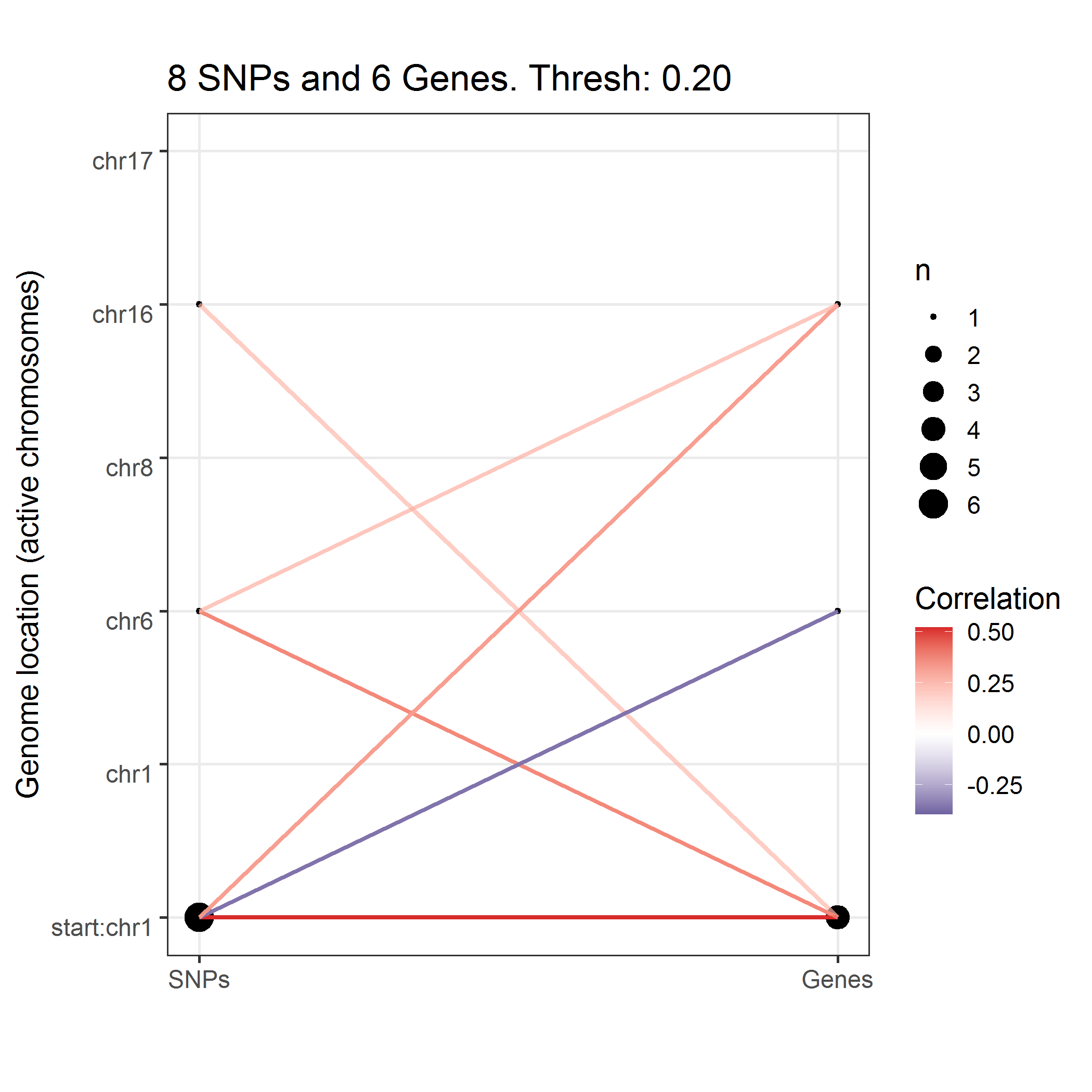} &
	    \includegraphics[width = \bmwidth]{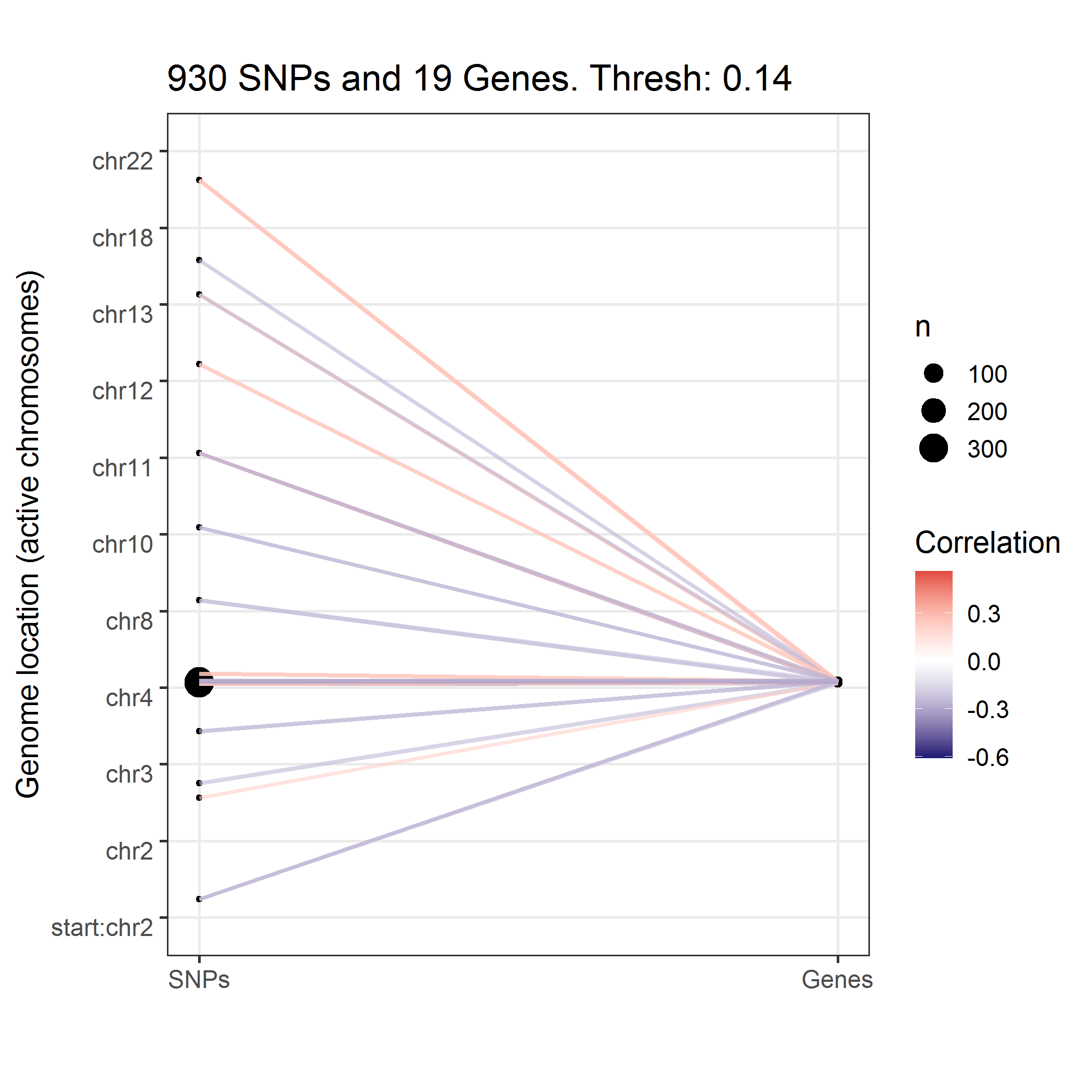}
	\end{tabular}
	\caption{Out of 31 BSP bimodules that had genes on 3 or more chromosomes and SNPs on 2 or more chromosomes, we selected 9 bimodules that looked interesting. The bipartite graph for each bimodule is formed out of the essential edges (Section \ref{sec:network-interpretation}).} 	\label{fig:bimod-plots-extra}
\end{figure}

See the plots in Figure \ref{fig:bimod-plots-extra}.
 
\subsection{Gene Ontology}
\label{sec:go-details}

Among the gene sets that we considered with $8$ or more genes, 18 out of the 145 BSP gene sets, and 1 out of the 5 CONDOR gene sets had significant overlap with GO categories.
Among the 40 GO terms detected by CONDOR, 27 terms were also found among the 135 terms detected by BSP. 
The complete list of the GO terms for the two methods now follows. We indicate statistical significance using * based on adjusted p-values. Notably, * is $10^{-2}$---$10^{-3}$, ** is $10^{-3}$---$10^{-5}$, *** is $10^{-5}$---$10^{-10}$, and **** is $< 10^{-10}$.

Significant GO terms for BSP:
\begin{longtable}[h]{rllr}
    \hline
   Signifier & GO.ID & Term & Bimodule \\ 
   \hline
   **** & GO:0060333 & interferon-gamma-mediated signaling path... &   1 \\ 
   **** & GO:0002478 & antigen processing and presentation of e... &   1 \\ 
   **** & GO:0019884 & antigen processing and presentation of e... &   1 \\ 
   **** & GO:0048002 & antigen processing and presentation of p... &   1 \\ 
   *** & GO:0019882 & antigen processing and presentation &   1 \\ 
   *** & GO:0071346 & cellular response to interferon-gamma &   1 \\ 
   *** & GO:0034341 & response to interferon-gamma &   1 \\ 
   *** & GO:0019886 & antigen processing and presentation of e... &   1 \\ 
   *** & GO:0002495 & antigen processing and presentation of p... &   1 \\ 
   *** & GO:0002504 & antigen processing and presentation of p... &   1 \\ 
   ** & GO:0045087 & innate immune response &   1 \\ 
   ** & GO:0050776 & regulation of immune response &   1 \\ 
   ** & GO:0006952 & defense response &   1 \\ 
   ** & GO:0031295 & T cell costimulation &   1 \\ 
   ** & GO:0031294 & lymphocyte costimulation &   1 \\ 
   * & GO:0050852 & T cell receptor signaling pathway &   1 \\ 
   * & GO:0002768 & immune response-regulating cell surface ... &   1 \\ 
   * & GO:0002764 & immune response-regulating signaling pat... &   1 \\ 
   * & GO:0050851 & antigen receptor-mediated signaling path... &   1 \\ 
   * & GO:0002682 & regulation of immune system process &   1 \\ 
   * & GO:0022409 & positive regulation of cell-cell adhesio... &   1 \\ 
   * & GO:0002253 & activation of immune response &   1 \\ 
   * & GO:0002429 & immune response-activating cell surface ... &   1 \\ 
   & GO:0006950 & response to stress &   1 \\ 
   & GO:0006955 & immune response &   1 \\ 
   & GO:0019221 & cytokine-mediated signaling pathway &   1 \\ 
   & GO:0002757 & immune response-activating signal transd... &   1 \\ 
   & GO:0050870 & positive regulation of T cell activation &   1 \\ 
   & GO:0002479 & antigen processing and presentation of e... &   1 \\ 
   & GO:1903039 & positive regulation of leukocyte cell-ce... &   1 \\ 
   & GO:0042590 & antigen processing and presentation of e... &   1 \\ 
   & GO:0045806 & negative regulation of endocytosis &   8 \\ 
   ** & GO:0050911 & detection of chemical stimulus involved ... &  11 \\ 
   ** & GO:0007608 & sensory perception of smell &  11 \\ 
   ** & GO:0050907 & detection of chemical stimulus involved ... &  11 \\ 
   * & GO:0009593 & detection of chemical stimulus &  11 \\ 
   * & GO:0007606 & sensory perception of chemical stimulus &  11 \\ 
   * & GO:0035459 & cargo loading into vesicle &  11 \\ 
   * & GO:0050906 & detection of stimulus involved in sensor... &  11 \\ 
   & GO:0000038 & very long-chain fatty acid metabolic pro... &  14 \\ 
   & GO:0006732 & coenzyme metabolic process &  14 \\ 
   & GO:0006417 & regulation of translation &  33 \\ 
   & GO:0034248 & regulation of cellular amide metabolic p... &  33 \\ 
   & GO:0010608 & posttranscriptional regulation of gene e... &  33 \\ 
   *** & GO:0046597 & negative regulation of viral entry into ... &  55 \\ 
   ** & GO:0035455 & response to interferon-alpha &  55 \\ 
   ** & GO:0035456 & response to interferon-beta &  55 \\ 
   ** & GO:0046596 & regulation of viral entry into host cell &  55 \\ 
   ** & GO:0045071 & negative regulation of viral genome repl... &  55 \\ 
   ** & GO:1903901 & negative regulation of viral life cycle &  55 \\ 
   ** & GO:0060337 & type I interferon signaling pathway &  55 \\ 
   ** & GO:0071357 & cellular response to type I interferon &  55 \\ 
   ** & GO:0034340 & response to type I interferon &  55 \\ 
   * & GO:0045069 & regulation of viral genome replication &  55 \\ 
   * & GO:0048525 & negative regulation of viral process &  55 \\ 
   * & GO:0019079 & viral genome replication &  55 \\ 
   * & GO:0046718 & viral entry into host cell &  55 \\ 
   * & GO:1903900 & regulation of viral life cycle &  55 \\ 
   * & GO:0030260 & entry into host cell &  55 \\ 
   * & GO:0044409 & entry into host &  55 \\ 
   * & GO:0051806 & entry into cell of other organism involv... &  55 \\ 
   * & GO:0051828 & entry into other organism involved in sy... &  55 \\ 
   * & GO:0043901 & negative regulation of multi-organism pr... &  55 \\ 
   & GO:0034341 & response to interferon-gamma &  55 \\ 
   & GO:0050792 & regulation of viral process &  55 \\ 
   & GO:0051607 & defense response to virus &  55 \\ 
   & GO:0051701 & interaction with host &  55 \\ 
   & GO:0043903 & regulation of symbiosis, encompassing mu... &  55 \\ 
   & GO:0009615 & response to virus &  55 \\ 
   *** & GO:0051225 & spindle assembly &  68 \\ 
   *** & GO:0007030 & Golgi organization &  68 \\ 
   *** & GO:0007051 & spindle organization &  68 \\ 
   ** & GO:0010256 & endomembrane system organization &  68 \\ 
   ** & GO:0000226 & microtubule cytoskeleton organization &  68 \\ 
   * & GO:0007017 & microtubule-based process &  68 \\ 
   * & GO:0070925 & organelle assembly &  68 \\ 
   & GO:0007010 & cytoskeleton organization &  68 \\ 
   **** & GO:0007156 & homophilic cell adhesion via plasma memb... &  70 \\ 
   **** & GO:0098742 & cell-cell adhesion via plasma-membrane a... &  70 \\ 
   *** & GO:0098609 & cell-cell adhesion &  70 \\ 
   ** & GO:0007155 & cell adhesion &  70 \\ 
   ** & GO:0022610 & biological adhesion &  70 \\ 
   * & GO:0007416 & synapse assembly &  70 \\ 
   * & GO:0007267 & cell-cell signaling &  70 \\ 
   * & GO:0006355 & regulation of transcription, DNA-templat... &  71 \\ 
   * & GO:1903506 & regulation of nucleic acid-templated tra... &  71 \\ 
   * & GO:2001141 & regulation of RNA biosynthetic process &  71 \\ 
   * & GO:0006351 & transcription, DNA-templated &  71 \\ 
   * & GO:0097659 & nucleic acid-templated transcription &  71 \\ 
   * & GO:0032774 & RNA biosynthetic process &  71 \\ 
   * & GO:0051252 & regulation of RNA metabolic process &  71 \\ 
   * & GO:2000112 & regulation of cellular macromolecule bio... &  71 \\ 
   * & GO:0010556 & regulation of macromolecule biosynthetic... &  71 \\ 
   * & GO:0019219 & regulation of nucleobase-containing comp... &  71 \\ 
   * & GO:0031326 & regulation of cellular biosynthetic proc... &  71 \\ 
   * & GO:0034654 & nucleobase-containing compound biosynthe... &  71 \\ 
   & GO:0009889 & regulation of biosynthetic process &  71 \\ 
   & GO:0018130 & heterocycle biosynthetic process &  71 \\ 
   & GO:0019438 & aromatic compound biosynthetic process &  71 \\ 
   & GO:0010468 & regulation of gene expression &  71 \\ 
   & GO:1901362 & organic cyclic compound biosynthetic pro... &  71 \\ 
   & GO:0016070 & RNA metabolic process &  71 \\ 
   **** & GO:0001580 & detection of chemical stimulus involved ... &  74 \\ 
   **** & GO:0050912 & detection of chemical stimulus involved ... &  74 \\ 
   **** & GO:0050913 & sensory perception of bitter taste &  74 \\ 
   **** & GO:0050909 & sensory perception of taste &  74 \\ 
   **** & GO:0050907 & detection of chemical stimulus involved ... &  74 \\ 
   **** & GO:0009593 & detection of chemical stimulus &  74 \\ 
   **** & GO:0007606 & sensory perception of chemical stimulus &  74 \\ 
   **** & GO:0050906 & detection of stimulus involved in sensor... &  74 \\ 
   *** & GO:0007600 & sensory perception &  74 \\ 
   *** & GO:0051606 & detection of stimulus &  74 \\ 
   *** & GO:0050877 & nervous system process &  74 \\ 
   ** & GO:0003008 & system process &  74 \\ 
   ** & GO:0007186 & G-protein coupled receptor signaling pat... &  74 \\ 
   * & GO:0006355 & regulation of transcription, DNA-templat... &  84 \\ 
   * & GO:1903506 & regulation of nucleic acid-templated tra... &  84 \\ 
   * & GO:2001141 & regulation of RNA biosynthetic process &  84 \\ 
   * & GO:0006351 & transcription, DNA-templated &  84 \\ 
   * & GO:0097659 & nucleic acid-templated transcription &  84 \\ 
   * & GO:0032774 & RNA biosynthetic process &  84 \\ 
   * & GO:0051252 & regulation of RNA metabolic process &  84 \\ 
   * & GO:2000112 & regulation of cellular macromolecule bio... &  84 \\ 
   * & GO:0010556 & regulation of macromolecule biosynthetic... &  84 \\ 
   * & GO:0019219 & regulation of nucleobase-containing comp... &  84 \\ 
   * & GO:0031326 & regulation of cellular biosynthetic proc... &  84 \\ 
   * & GO:0034654 & nucleobase-containing compound biosynthe... &  84 \\ 
   & GO:0009889 & regulation of biosynthetic process &  84 \\ 
   & GO:0018130 & heterocycle biosynthetic process &  84 \\ 
   & GO:0019438 & aromatic compound biosynthetic process &  84 \\ 
   & GO:0010468 & regulation of gene expression &  84 \\ 
   & GO:1901362 & organic cyclic compound biosynthetic pro... &  84 \\ 
   & GO:0016070 & RNA metabolic process &  84 \\ 
   *** & GO:1901685 & glutathione derivative metabolic process &  93 \\ 
   *** & GO:1901687 & glutathione derivative biosynthetic proc... &  93 \\ 
   ** & GO:0006749 & glutathione metabolic process &  93 \\ 
   ** & GO:0042178 & xenobiotic catabolic process &  93 \\ 
   ** & GO:0042537 & benzene-containing compound metabolic pr... &  93 \\ 
   * & GO:0006575 & cellular modified amino acid metabolic p... &  93 \\ 
   * & GO:0044272 & sulfur compound biosynthetic process &  93 \\ 
   * & GO:0046854 & phosphatidylinositol phosphorylation &  95 \\ 
   * & GO:0046834 & lipid phosphorylation &  95 \\ 
   & GO:0048015 & phosphatidylinositol-mediated signaling &  95 \\ 
   & GO:0048017 & inositol lipid-mediated signaling &  95 \\ 
   ** & GO:0006882 & cellular zinc ion homeostasis & 113 \\ 
   ** & GO:0055069 & zinc ion homeostasis & 113 \\ 
   ** & GO:0010273 & detoxification of copper ion & 113 \\ 
   ** & GO:1990169 & stress response to copper ion & 113 \\ 
   * & GO:0061687 & detoxification of inorganic compound & 113 \\ 
   * & GO:0097501 & stress response to metal ion & 113 \\ 
   * & GO:0071294 & cellular response to zinc ion & 113 \\ 
   * & GO:0071280 & cellular response to copper ion & 113 \\ 
   & GO:0046916 & cellular transition metal ion homeostasi... & 113 \\ 
   & GO:0071276 & cellular response to cadmium ion & 113 \\ 
   & GO:0046688 & response to copper ion & 113 \\ 
   & GO:0055076 & transition metal ion homeostasis & 113 \\ 
   & GO:0072488 & ammonium transmembrane transport & 137 \\ 
   & GO:0006089 & lactate metabolic process & 139 \\ 
   & GO:0006882 & cellular zinc ion homeostasis & 142 \\ 
   & GO:0055069 & zinc ion homeostasis & 142 \\ 
   & GO:0006882 & cellular zinc ion homeostasis & 143 \\ 
   & GO:0055069 & zinc ion homeostasis & 143 \\ 
   \hline
\end{longtable}

Significant GO terms for CONDOR
\begin{longtable}[h]{rllr}
  \hline
 Signifier & GO.ID & Term & Bimodule \\ 
 \hline
 * & GO:0050852 & T cell receptor signaling pathway &   1 \\ 
 * & GO:0050851 & antigen receptor-mediated signaling path... &   1 \\ 
 & GO:0006355 & regulation of transcription, DNA-templat... &   1 \\ 
 & GO:1903506 & regulation of nucleic acid-templated tra... &   1 \\ 
 & GO:2001141 & regulation of RNA biosynthetic process &   1 \\ 
 * & GO:0060333 & interferon-gamma-mediated signaling path... &   2 \\ 
 **** & GO:0002478 & antigen processing and presentation of e... &   4 \\ 
 **** & GO:0019884 & antigen processing and presentation of e... &   4 \\ 
 **** & GO:0048002 & antigen processing and presentation of p... &   4 \\ 
 **** & GO:0019886 & antigen processing and presentation of e... &   4 \\ 
 *** & GO:0002495 & antigen processing and presentation of p... &   4 \\ 
 *** & GO:0002504 & antigen processing and presentation of p... &   4 \\ 
 *** & GO:0019882 & antigen processing and presentation &   4 \\ 
 *** & GO:0031295 & T cell costimulation &   4 \\ 
 *** & GO:0031294 & lymphocyte costimulation &   4 \\ 
 *** & GO:0060333 & interferon-gamma-mediated signaling path... &   4 \\ 
 ** & GO:0050852 & T cell receptor signaling pathway &   4 \\ 
 ** & GO:0050870 & positive regulation of T cell activation &   4 \\ 
 ** & GO:1903039 & positive regulation of leukocyte cell-ce... &   4 \\ 
 ** & GO:0050778 & positive regulation of immune response &   4 \\ 
 ** & GO:0002253 & activation of immune response &   4 \\ 
 ** & GO:0050851 & antigen receptor-mediated signaling path... &   4 \\ 
 ** & GO:0022409 & positive regulation of cell-cell adhesio... &   4 \\ 
 ** & GO:0071346 & cellular response to interferon-gamma &   4 \\ 
 ** & GO:0051251 & positive regulation of lymphocyte activa... &   4 \\ 
 * & GO:1903037 & regulation of leukocyte cell-cell adhesi... &   4 \\ 
 * & GO:0034341 & response to interferon-gamma &   4 \\ 
 * & GO:0002696 & positive regulation of leukocyte activat... &   4 \\ 
 * & GO:0050863 & regulation of T cell activation &   4 \\ 
 * & GO:0050867 & positive regulation of cell activation &   4 \\ 
 * & GO:0007159 & leukocyte cell-cell adhesion &   4 \\ 
 * & GO:0050776 & regulation of immune response &   4 \\ 
 * & GO:0002429 & immune response-activating cell surface ... &   4 \\ 
 * & GO:0002684 & positive regulation of immune system pro... &   4 \\ 
 * & GO:0006955 & immune response &   4 \\ 
 * & GO:0022407 & regulation of cell-cell adhesion &   4 \\ 
 & GO:0045087 & innate immune response &   4 \\ 
 & GO:0002768 & immune response-regulating cell surface ... &   4 \\ 
 & GO:0045785 & positive regulation of cell adhesion &   4 \\ 
 & GO:0002455 & humoral immune response mediated by circ... &   4 \\ 
 & GO:0051249 & regulation of lymphocyte activation &   4 \\ 
 & GO:0019221 & cytokine-mediated signaling pathway &   4 \\ 
 & GO:0042110 & T cell activation &   4 \\ 
 \hline
\end{longtable}

\section{Application of BSP to North American Temperature and Precipitation Data}
\label{sec:climate}

\subsection{Introduction} 
 
The relationship between temperature and precipitation over North America has been well 
documented \citep{madden_williams_1978, berg2015, adler2008, livneh2016,HAO2018} and is 
of agricultural importance. 
For example, \cite{berg2015} noted widespread correlation between summertime mean temperature and precipitation  at the same location over various land regions. We explore these relationships using the Bimodule Search Procedure. In particular, the method allows us to search for clusters of distal temperature-precipitation relationships, known as teleconnections, whereas previous work has mostly focused on analyzing spatially proximal correlations.

 We applied BSP to find pairs of geographic regions such that summer temperature 
in the first region is significantly correlated in aggregate with summer precipitation in the second region one year later. 
We will refer to such region pairs as T-P (temperature-precipitation) bimodules.
T-P bimodules reflect mesoscale analysis of region-specific climatic patterns, which can be useful for predicting impact of climatic changes on practical outcomes like agricultural output.

\subsection{Data Description and Processing} 

The Climatic Research Unit (CRU TS version 4.01) data \citep{CRU_paper} contains daily global measurements 
of temperature (T) and precipitation (P) levels on land over a $.5^\text{o}  \times .5^\text{o}$ 
(360 pixels by 720 pixels) resolution grid from 1901 to 2016. 
We reduced the resolution of the data to $2.5^\text{o} \times 2.5^\text{o}$ (72 by 144 pixels) by averaging over neighboring pixels and
 restricted to 427 pixels corresponding to the latitude-longitude pairs 
within North America.
For each available year and each pixel/location we averaged temperature (T) and precipitation (P) over the summer months 
of June, July, and August. Each feature of the resulting time series was centered and scaled to have zero mean and unit variance. 
The data matrix $\mathbb{X}$, reflecting temperature, had 115 rows containing the annual summer-aggregated temperatures 
from 1901 to 2015 for each of the 427 locations. 
The data matrix $\mathbb{Y}$, reflecting precipitation, had 115 rows containing the annual summer-aggregated precipitation 
from 1902 to 2016 (lagged by one year from temperature) for each of the 427 locations.

Analysis of summer precipitation versus summer temperatures lagged by 2 years, and temperatures from different seasons  (winter T; summer P of the same year)
in the same year did not yield any bimodules.

\subsection{Bimodules Search Procedure and Diagnostics} \label{sec:BSP_Climate}

We applied BSP on the climate data with the false discovery parameter $\alpha=0.045$,
selected using the procedure in Section \ref{sec:howtochoosealpha}, to keep edge-error 
under 0.1 (see Figure \ref{fig:lineplots}, \smref{sec:climate-appendix}). 
BSP searches for groups of temperature and precipitation pixels that have significant aggregate cross-correlation.
Temperature and precipitation data exhibit both spatial and temporal auto-correlations. 
The BSP procedure does not make use of the pixel locations.  While the permutation null employed by BSP 
directly accounts for spatial-correlations within the temperature and the precipitation data, we note that it does not 
directly account for temporal correlations, which violate a sample exchangeability assumption used in the 
p-value approximations. The temporal auto-correlation in our data was moderate, ranging from 0.10 to 0.30 
for various features.

BSP found five distinct bimodules; the effective number of bimodules was three. 
After filtering, the two bimodules illustrated in Figure \ref{fig: CRU_offset} and another bimodule with 80 temperature pixels and 5 precipitation pixels remained.  We omitted a further analysis of the third bimodule as 
its precipitation pixels were same as those of Bimodule $2$ in Figure \ref{fig: CRU_offset} and 
its temperature pixels were geographically scattered.

 \newcommand\y{.33}
   \begin{figure}[tbp]  
   	 \centering
   	       \textbf{CRU: T(JJA)-P(JJA, offset), 1901-2016}, $\alpha = .045 $
 	\begin{tabular}{ccc}  
 \includegraphics[ width = \y \linewidth, 
   	 trim={2.2cm, 2.1cm, 3cm, 2.6cm}, clip]{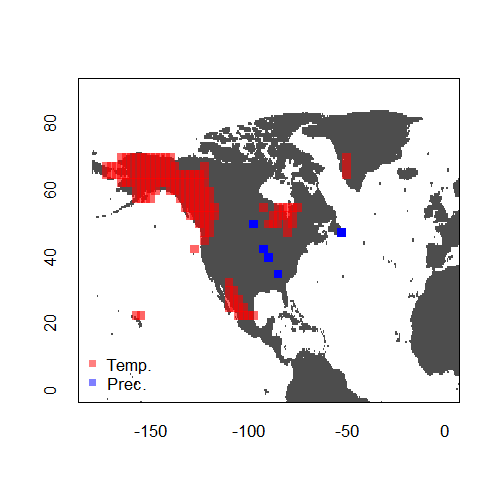}
&
   	 	\includegraphics[ width = \y \linewidth,  
   	 trim={2.2cm, 2.1cm, 3cm, 2.6cm}, clip]{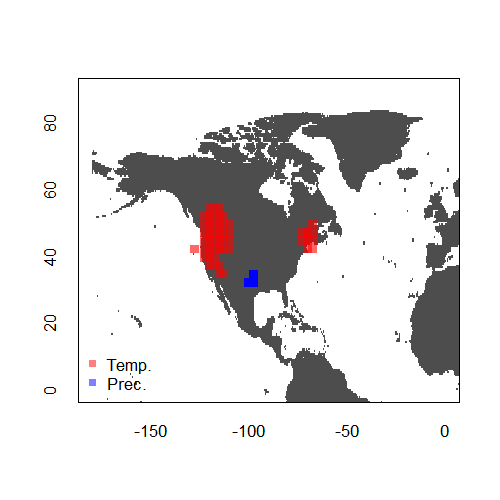}
   	 \end{tabular} 
         \caption{\footnotesize{}  Bimodules of summer temperature and precipitation in North America from CRU observations from 1901-2016. The left bimodule ($1$) contains 149 temperature locations (pixels) and 6 precipitation locations. The right bimodule ($2$) contains 53 temperature and 5 precipitation locations.}
   	\label{fig: CRU_offset} 
   \end{figure}

Temperature pixels in Bimodules 1 and 2 are situated distally from the precipitation pixels, 
but the temperature and precipitation pixels within each bimodule form blocks of contiguous geographical regions.    
Since BSP did not make use of any location information when searching for bimodules,  
these effects might have a common spatial origin.
      
The locations regions identified by the bimodules occupy large geographical areas.
The precipitation pixels from Bimodule 1 form a vertical stretch around the eastern edge of the Great Plains and are correlated with temperature pixels in large areas of the Pacific Northwest, Alaska, and Mexico. 
In Bimodule 2 precipitation in the southern Great Plains around Oklahoma is strongly correlated with temperature 
in the Northwestern Great Plains. 
An anomalously hot summer Oregon in one year in the Northwest 
suggests an anomalously rainy growing season in the following year in the Southern Great Plains. 
Pixel-wise positive correlations are discussed in \smref{sec:climate-appendix}.  
The coastal proximity of all the temperature clusters suggest influences of oscillations in sea surface temperatures.  Aforementioned patterns from both bimodules map to locations of agricultural productivity, such as in Oklahoma and Missouri (Figure \ref{fig: CRU_offset}). 
       
The bimodules found by BSP only consider the magnitudes of correlations between 
temperature and precipitation.   Further analysis of these bimodules shows that the 
significant correlations between temperature and precipitation are positive in the Great Plains region.
These results agree with findings on concurrent T-P correlations in the Great Plains \citep{zhao_khalil, berg2015,Wang22512}, which noted widespread correlations between summertime mean temperatures and precipitation at the same location over land in various parts of North America, notably the Great plains. Our findings  show strong correlations between northwestern (coastal) temperatures and Great Plains precipitation and generally agree with findings in the literature. For example, 
 \cite{livneh2016} considered the relationship between hot temperatures and droughts in the Great Plains, noting that hot temperatures in the summer are related to droughts in the following year on the overall global scale. 
The results of  \cite{livneh2016} preface the  results contained within the above bimodules, but the latter are additionally able to find regions where this effect is significant.

   Our findings  demonstrate the utility of BSP   in finding  insights   into remote  correlations   {between precipitation and temperature} in North America.
   Further research may build on these exploratory findings and create a model that can forecast precipitation in agriculturally productive regions around the world.

\section{Climate Analysis Details}
\label{sec:climate-appendix}

\begin{figure}\centering
  	\begin{tabular}{cc}
  				\begin{turn}{90}
  			\quad \quad \quad \quad    Edge-error estimate  
  		\end{turn}   &  \includegraphics[width = .45\linewidth,	trim={1cm 1cm 0cm 0cm}, clip ]{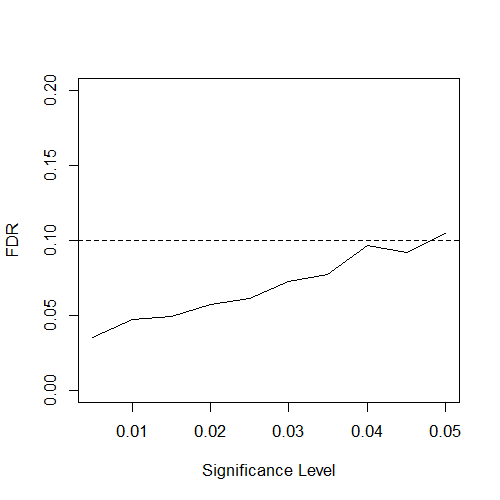} \\
  		&      False discovery parameter ($\alpha$)
  	\end{tabular}
 
 \caption{\footnotesize{} Average edge-error estimates for BSP results for the climate data based on $100$ half-permutations (Section \ref{sec:half-perm}) for $\alpha$ ranging from 0.01 to 0.05.  The edge-error estimates exceed $0.05$ for the first time at $\alpha=0.045$.}
\label{fig:lineplots} 
\end{figure}

  Figure \ref{fig:lineplots} shows the edge-error estimates we used to choose $\alpha$. Table \ref{tab:bimod-cross-summary} shows a summary of cross-correlations for each precipitation pixel from the two BSP bimodules.
  
\begin{table}\centering
$A$
	\begin{tabular}{rrr}
		\hline
			$P$ 	Pixel & Mean & SD \\ 
		\hline
		1 & 0.28 & 0.07 \\ 
		2 & 0.27 & 0.06 \\ 
		3 & 0.28 & 0.08 \\ 
		4 & 0.27 & 0.08 \\ 
		5 & 0.31 & 0.06 \\ 
		6 & 0.30 & 0.08 \\ 
		\hline
	\end{tabular}

	\centering
$B$
	\begin{tabular}{rrr}
		\hline
		$P$ Pixel & Mean & SD \\ 
		\hline
		1 & 0.31 & 0.04 \\ 
		2 & 0.35 & 0.03 \\ 
		3 & 0.29 & 0.04 \\ 
		\hline
	\end{tabular}
\caption{Summary of the cross-correlations for each precipitation ($P$) pixel in the two BSP bimodules $A$ and $B$ from the climate data. Each entry shows the mean and standard deviation of the cross-correlations of each $P$ in the bimodule with other $T$ pixels in the same bimodule. Results show that all the cross-correlations tend to be strong and positive.} \label{tab:bimod-cross-summary}
\end{table}
 
\clearpage
\bibliographystyle{plainnat}
\bibliography{refs}
\end{document}